\documentclass[prb,%
 reprint,
 amsmath,amssymb,
 aps,
]{revtex4-2}

\usepackage{graphicx}

\usepackage{dcolumn}
\usepackage{bm}
\usepackage{xcolor}
\usepackage{hyperref}

\newcommand{\R}{\mathbb{R}} 
\newcommand{\C}{\mathbb{C}}
\newcommand{\Z}{\mathbb{Z}}

\newcommand{\Ha}{\mathcal{H}}
\newcommand{\dd}{\mathrm{d}}
\newcommand{\ep}{\epsilon}

\newcommand{\la}{\langle}
\newcommand{\ra}{\rangle}
\newcommand{\di}{\partial}

\newcommand{\g}{\gamma}
\newcommand{\T}{{\mathcal{T}}}
\newcommand{\Oo}{{\mathcal{O}}}
\newcommand{\G}{{\mathcal{G}}}
\newcommand{\up}{{\uparrow}}
\newcommand{\dow}{{\downarrow}}
\newcommand{\half}{{\frac{1}{2}}}
\newcommand{\sqth}{{\frac{\sqrt{3}}{2}}}
\newcommand{\Aa}{{\mathcal{A}}}
\newcommand{\rr}{{{\boldsymbol{r}}}}
\newcommand{\kk}{{{\boldsymbol{k}}}}
\newcommand{\qq}{{{\boldsymbol{q}}}}
\newcommand{\ee}{{\boldsymbol{\hat{e}}}}
\newcommand{\ncs}{{n_{\overline{C}_6S}}}
\newcommand{\nst}{{n_{ST_1}}}
\newcommand{\nc}{{n_{\overline{C}_6}}}
\newcommand{\przo}{\textrm{Pr}_2 \textrm{Zr}_2 \textrm{O}_7}
\newcommand{\prhfo}{\textrm{Pr}_2 \textrm{Hf}_2 \textrm{O}_7}
\newcommand{\ze}{\mathbf{0}}

\begin{document}

\preprint{APS/123-QED}

\title{Quantum Spin Liquids in Pyrochlore Magnets With Non-Kramers Local Moments}
\author{Tony An}
 \email{t.an@mail.utoronto.ca}
\author{F\'elix Desrochers}%
\author{Yong Baek Kim}%
\affiliation{
 Department of Physics, University of Toronto, Toronto, Ontario M5S 1A7, Canada
}

\date{\today}

\begin{abstract}
Numerous experiments on pyrochlore oxides Pr$_2$(Zr, Sn, Hf, Ir)$_2$O$_7$ with non-Kramers Pr$^{3+}$ ions suggest that they support a quantum spin liquid (QSL) ground state, but the precise nature of the QSL remains unclear. Quantum spin ice with dominant dipolar Ising and smaller quadrupolar transverse exchange interactions is one such candidate, but a dominant inelastic neutron scattering signal suggests that such a picture may not be consistent with experimental results. The microscopic exchange couplings of these compounds are also not known, leaving room for many possible QSL states. In this work, we use Schwinger boson mean-field theory supplemented by a projective symmetry group classification to study possible $\Z_2$ QSLs in pyrochlore magnets with dipolar-quadrupolar non-Kramers local moments. We build a mean-field phase diagram and find four QSLs in the frustrated region of parameter space that are consistent with inelastic signals observed in neutron scattering data on Pr$_2$Zr$_2$O$_7$ and Pr$_2$Hf$_2$O$_7$. Among these, two robust QSLs occur in the regime with dominant transverse exchange rather than Ising exchange. We then compute the static and dynamic spin structure factors for these QSL candidates, which can be used to distinguish them in neutron scattering experiments.
\end{abstract}

\maketitle


\section{\label{sec:level1}Introduction}
Quantum spin liquids (QSLs) are novel quantum paramagnetic states of matter produced by long-range entanglement of interacting spins \cite{Savary2016, wen2017colloquium, knolle2019field, zhou2017quantum, Broholm2020}. QSLs do not order at zero temperature, and they possess fractionalized excitations such as spinons and emergent gauge fields \cite{Ioffe1989, wen1990topological, wen2004quantum, levin2005colloquium, chen2010local, wen2013topological, wen2017colloquium, wen2019choreographed}. While a dimer liquid form of QSL is realized in Rydberg atom arrays \cite{Verresen2021, Semeghini2021}, the discovery of QSL in real materials remains an outstanding challenge in quantum materials research.

\begin{figure}
    \centering
    \includegraphics[width=\linewidth]{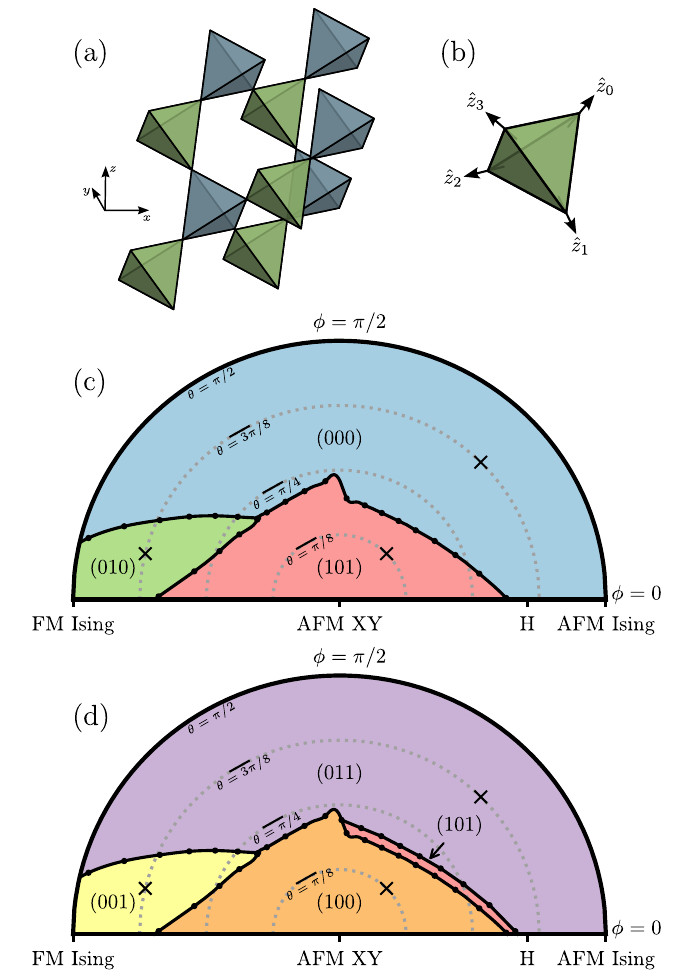}
    \caption{\label{fig:fig 1}(a) The pyrochlore lattice and global Cartesian coordinate frame, where ``up-pointing" and ``down-pointing" tetrahedra are coloured green and blue, respectively. (b) An up-pointing tetrahedron with local $\hat{z}$ axes and sublattice indices attached. (c) Phase diagram in the region $J_\pm < 0$ with $J_{zz}^2+J_\pm^2+J_{\pm\pm}^2=1$ indicating the lowest energy PSG classes. We set $\kappa=0.1$. The XXZ model occurs along the horizontal line joining $\phi=0$ to $\phi=\pi$. The point H indicates the antiferromagnetic Heisenberg point $J_\pm/J_{zz}=-\frac{1}{2}, J_{\pm\pm}=0$. The $\times$'s indicate the points at which the structure factors were calculated for the relevant PSG classes (see table \ref{table:params}). (d) Phase diagram at $\kappa=0.1$ indicating the second-lowest energy PSG classes.}
\end{figure}

Pyrochlore magnets have been considered as possible platforms for QSLs due to the geometric frustration of the pyrochlore lattice (Fig.~\ref{fig:fig 1} (a)) \cite{hermele2004pyrochlore, huse2003coulomb, jaubert2011analysis, banerjee2008unusual, shannon2012quantum, Onoda2011, huang2018dynamics, huang2020extended, henley2010coulomb, bramwell2020history, castelnovo2012spin, Gardner2010, Book2011, udagawa2021spin, Rau2019}. In particular, several promising signatures of QSL behaviour have been observed in pyrochlore rare-earth oxides $\textrm{Pr}_2(\textrm{Zr, Sn, Hf, Ir})_2\textrm{O}_7$ \cite{Kimura2013, Petit2016, Martin2017, Zhou2008, Ortiz2024, hatnean2016single, Sibille2016, Sibille2018, Machida2005, Nakatsuji2006}. These compounds have the interesting property that the magnetically active $\textrm{Pr}^{3+}$ ions are described as a non-Kramers doublet at low energies, which is protected by crystal symmetries, but not by Kramers degeneracy \cite{Rau2019}. The non-Kramers doublet is conveniently represented by pseudospins-$1/2$, where the $S^z$ component transforms as a dipole but the local $S^x$ and $S^y$ components transform as quadrupoles \cite{Rau2019, Curnoe2018}. Heat capacity measurements down to $0.35$ K and DC magnetic susceptibility measurements down to $1.8$ K show no evidence of a phase transition \cite{Kimura2013,Petit2016,Zhou2008,Ortiz2024, Sibille2016,Nakatsuji2006}. Furthermore, neutron scattering experiments in $\przo$, $\prhfo$, and $\textrm{Pr}_2\textrm{Sn}_2\textrm{O}_7$ find an inelastic continuum of excitations that could be attributed to fractionalized quasiparticles of a QSL \cite{Kimura2013, Petit2016, Wen2017, Sibille2018, Zhou2008, Ortiz2024}. 

It has been suggested that the putative QSL in $\przo$ and $\prhfo$ is quantum spin ice (QSI) \cite{Kimura2013, Tokiwa2018, Sibille2018} with dominant dipolar Ising exchange and smaller quadrupolar transverse exchange interactions. The minimal model for QSI is the XXZ model on the pyrochlore lattice 
\begin{equation} \label{eq:xxz_model}
    H_\textrm{XXZ} = \sum_{\la ij\ra} \left[ J_{zz} S_i^z S_j^z - J_\pm (S_i^+ S_j^- + S_i^- S_j^+) \right],
\end{equation} 
where $J_{zz}>0$. In classical spin ice (i.e., $J_\pm=0$), the ground state spin configuration is characterized by the local two-in-two-out constraint, resulting in a macroscopic degeneracy \cite{Gingras2014}. An excited state with energy gap $J_{zz}$ can be produced by flipping one spin in a particular ground state, creating two ``defect tetrahedra" which break the two-in-two-out constraint \cite{Gingras2014, bramwell2001spin, castelnovo2008magnetic}. Going to the quantum theory in the perturbative regime $J_{\pm}\ll J_{zz}$, the action of the transverse operators $S^\pm$ on the ground state is to create a pair of dispersive spinon-antispinon excitations with an excitation gap on the order of $J_{zz}/2$. Formally, QSI can be described by mapping pseudospins to lattice quantum electrodynamics~\cite{hermele2004pyrochlore, banerjee2008unusual, Benton2012, Savary2012, Lee2012}
\begin{subequations}    
\begin{align}
    S^z_i &\sim \mathcal{E}_{rr'} \\
    S^\pm_i &\sim \Phi^\dagger_r \Phi_{r'},
\end{align} 
\end{subequations}
where $i$ labels the sites of the pyrochlore lattice and $r,r'$ represent the centers of the tetrahedra sharing the site $i$. Here, $\mathcal{E}_{rr'}$ is a lattice electric field and $\Phi^\dagger_r \Phi_{r'}$ represents the creation of a spinon and antispinon pair. Physically, the spinons (``electric charges'') are the quantum analog of the classical defect tetrahedra, while fluctuations of the electric field describe gapless photon excitations of the QSI \cite{hermele2004pyrochlore, banerjee2008unusual, Benton2012, Savary2012, Lee2012, savary2021quantum}. 

In the case of Pr-based pyrochlore magnets, if the ground state is a QSI with the dominant dipolar Ising exchange interaction, only the dipolar Ising moment $S^z$ or the emergent electric field couples linearly to the neutron spin. On the other hand, the gapped excitations of QSI involve $S^\pm$ quadrupolar moments and hence they do not couple to neutron spins. Furthermore, although the dual magnetic monopoles or ``vison'' excitations can couple to neutrons \emph{in principle}~\cite{chen2017dirac}, detailed calculations show that such a signal is several orders of magnitude weaker than any contribution from the photons~\cite{kwasigroch2020vison}. Hence, neutron scattering is expected to detect the quasielastic low-energy photons while being essentially insensitive to the inelastic signals from gapped spinon excitations and the extremely weak magnetic monopole contribution. On the other hand, neutron scattering in $\przo$ and $\textrm{Pr}_2\textrm{Hf}_2\textrm{O}_7$ shows significant spectral weight in the inelastic signals, which is inconsistent with the interpretation of the QSI \cite{Kimura2013, Petit2016, Wen2017, anand2016physical, Sibille2018}. Interestingly, the inelastic signals show ``rod"-like structures in the $[hhl]$ scattering plane~\cite{Castelnovo2019}. It has been suggested that these inelastic signals may be due to disorder effects \cite{Bonville2016,Martin2017,Wen2017,Savary2017,Benton2018}. Finally, there is no consensus on the exchange parameters of $\textrm{Pr}_2(\textrm{Zr, Sn, Hf, Ir})_2\textrm{O}_7$, so we have no clear reason to assume that they lie in or even near the spin ice regime in parameter space even without disorder.  

Given these facts, we explore alternative quantum spin liquid candidates in pyrochlore magnets with non-Kramers local moments. In this work, instead of considering disorder effects, we investigate the feasibility of quantum spin liquids beyond the quantum spin ice regime, which may explain the dominant inelastic signals in neutron scattering experiments. We study possible $\Z_2$ QSLs using Schwinger boson mean-field theory, restricting our attention to QSLs which do not break any space group symmetries or time-reversal symmetry. We use the projective symmetry group (PSG) to classify all such QSLs~\cite{Wen2002, Wang2006, Liu2019}. We construct the phase diagram and identify the spin liquid candidates that are consistent with the experiments. Among them, the most robust candidates occur in the region where the transverse quadrupolar exchange coupling $J_\pm$ is dominant. 
This is a parameter regime where studies on the XXZ model~\eqref{eq:xxz_model} have identified a $U(1)$ nematic QSL through exact diagonalization and variational calculations~\cite{benton2018quantum} and a disordered state with a large nematic susceptibility in pseudo-Majorana fermion functional renormalization group~\cite{Schaden2025}. The $\Z_2$ QSLs identified here might be closely competing with these disordered states and may be stabilized by couplings beyond the pristine XXZ model. Finally, we compute the static and dynamic spin structure factors for these candidates to identify their experimental signatures in neutron scattering. 

\section{Model}
In praseodymium-based rare-earth pyrochlores, magnetic $\textrm{Pr}^{3+}$ ions are located at the shared vertices of neighbouring tetrahedra \cite{Gardner2010}. Each $\textrm{Pr}^{3+}$ has two $f$-electrons, and an application of Hund's rules results in a degenerate $J=4$ total angular momentum state. A crystal electric field (CEF) with $D_{3d}$ symmetry splits the $J=4$ manifold into doublets and singlets. Since there are an even number of electrons, Kramers' theorem does not apply, so any degeneracy arises from crystal symmetries \cite{Rau2019}. Experimentally, it is found that the lowest-lying states form a doublet (i.e., a non-Kramers doublet) and are well separated from the higher CEF states by at least $9$ meV \cite{Machida2005, Martin2017, Sibille2018}. Thus, we are justified in keeping only the non-Kramers doublet degrees of freedom for a low-energy description of the physics. We denote the two degenerate states by $|\pm\ra$ and define pseudospin-$1/2$ operators $S$ by 
\begin{subequations}
    \begin{align}
        S^z &= \frac{1}{2} \Big(|+\ra\la+| - |-\ra\la-|\Big) \\
        S^\pm &= |\pm\ra\la\mp|.
    \end{align}    
\end{subequations}
Explicitly, the doublet can be written in terms of the standard angular momentum basis $|J,m\ra$ with $J=4$: 
\begin{align}
    |\pm\ra &= \alpha |m=\pm 4\ra \pm \beta|m=\pm1\ra -\g |m=\mp 2\ra \label{eq:doublet}
\end{align} where $\alpha,\beta,\g \in \R$ \cite{Schaffer2013}. The resulting $S^z$ component of the non-Kramers doublet transforms as a dipole, while the in-plane $S^x, S^y$ components transform as quadrupoles \cite{Onoda2011}. 

The pyrochlore lattice is a corner-sharing network of tetrahedra with four sublattices labeled by $\mu \in \{0,1,2,3\}$, shown in Fig. \ref{fig:fig 1} (a). The centers of the tetrahedra form a diamond lattice. In global Cartesian coordinates (GCC), we specify the position of the unit cell with the basis vectors 
\begin{subequations}
\begin{align}
    \ee_1 &= \frac{1}{2}(0,1,1)_\textrm{global}\\
    \ee_2 &= \frac{1}{2}(1,0,1)_\textrm{global}\\
    \ee_3 &= \frac{1}{2}(1,1,0)_\textrm{global}.
\end{align}     
\end{subequations}
For notational convenience, we define $\ee_0 = (0,0,0)_\textrm{global}$. We can specify the lattice sites by adding $\hat{\ep}_\mu = \frac{1}{2}\ee_\mu$, defined as the displacement of the $\mu$ sublattice from the $0^{\mathrm{th}}$ sublattice, to the unit cell position. Thus, the sublattice-indexed pyrochlore coordinates are given by 
\begin{align}
    \rr_\mu &= r_1 \ee_1 + r_2\ee_2 + r_3 \ee_3 + \hat{\ep}_\mu \\ 
    &= \frac{1}{2}(r_2+r_3,r_1+r_3, r_1+r_2)_\textrm{global} + \frac{1}{2}\ee_\mu \label{eq:sipc}
\end{align}

The space group of the pyrochlore lattice is generated by the following operators: 
\begin{itemize}
    \item $T_i$: translation by $\ee_i$ where $i\in \{1,2,3\}$.
    \item $\overline{C}_6$: threefold rotation about the $[111]$ axis ($\ee_1+\ee_2+\ee_3$) composed with inversion.
    \item $S$: a ``screw" operation defined by fractional translation by $\hat{\ep}_3 = \frac{1}{2}\ee_3$ composed with a $\pi$ rotation around $\ee_3$. \cite{ Curnoe2018, Liu2019, Desrochers2022}
\end{itemize}

Finally, it is convenient to work in sublattice-dependent local bases to express the orientations of the pseudospins. We attach an orthonormal basis $\{\hat{x}_\mu, \hat{y}_\mu, \hat{z}_\mu\}$ to each sublattice $\mu$ (expressions given in appendix \ref{appmicroscopic}). As shown in Fig. \ref{fig:fig 1} (b), these have the property that each $\hat{z}_\mu$ basis vector points out of the ``up-pointing" tetrahedra.

Written in the sublattice-dependent local bases, the most general symmetry-allowed nearest-neighbour Hamiltonian for the non-Kramers doublet is 
\begin{align}
    H &= \sum_{\la ij\ra} \Big[ J_{zz} S^z_i S^z_j - J_\pm (S^+_i S^-_j + S^-_i S^+_j) \nonumber \\ 
    &+ J_{\pm\pm} (\g_{ij} S^+_i S^+_j + \g_{ij}^* S^-_i S^-_j) \Big],\label{eq:pseudospinham}
\end{align} where $i,j$ denote pyrochlore lattice sites, and $\g_{ij}$ is a bond-dependent phase factor defined in appendix \ref{appmicroscopic} \cite{Onoda2011}. We take $J_{\pm\pm} > 0$ without loss of generality because its sign can be flipped by a $\frac{\pi}{2}$ rotation of each local basis $\{\hat{x}_\mu, \hat{y}_\mu, \hat{z}_\mu\}$ about $\hat{z}_\mu$ \cite{Chung2024}.

\section{Schwinger Boson Mean-Field Theory and Classification of Spin Liquids}
\subsection{Schwinger Bosons}

We describe prospective QSLs using the Schwinger boson construction, in which the pseudospins \eqref{eq:pseudospinham} at pyrochlore site $i$ are expressed as
\begin{align}
    S^a_i &= \frac{1}{2}\sum_{\alpha\beta} b^\dagger_{i\alpha} [\sigma^a]_{\alpha\beta} b_{i\beta} \label{eq:schwingerboson}
\end{align} 
where $a \in \{x,y,z\}$ and $\sigma^a$ are the Pauli matrices. The operators $b_{i\alpha}$ satisfy the bosonic commutation relations $[b_{i\alpha}, b^\dagger_{j\beta}] = \delta_{ij}\delta_{\alpha\beta}$. The Schwinger bosons physically represent the fractionalized excitations of the QSL, which we refer to as spinons. If we impose the local constraint \begin{equation}
    n_{i} = \sum_\alpha b^\dagger_{i\alpha} b_{i\alpha} = \kappa
\end{equation} with $\kappa = 1$, we obtain a faithful representation of the original pseudospin-$1/2$ Hilbert space \cite{Book2011, Wang2006}. In the mean-field theory, this constraint is relaxed to only be satisfied on average
\begin{equation}
    \la n_{i} \ra = \kappa \label{eq:occconstraint}
\end{equation} 
for all sites $i$. We will also treat $\kappa$ as a continuous positive parameter, which gives us control over the size of quantum effects. Taking $\kappa \to \infty$ recovers the classical limit, and $\kappa < 1 $ corresponds to larger quantum fluctuations. In the current work, we focus on small $\kappa \ll 1$ to obtain all the competing QSLs, and comment on what is expected in the physical $\kappa=1$ limit later in sections \ref{phase properties} and \ref{discussion} \cite{Messio2010}. As $\kappa$ increases, magnetic order may develop in some parts of the phase diagram, signaled by the closing of the spinon gap and condensation of Schwinger bosons \cite{Wang2006, Dey2024, Liu2019}. 

\subsection{Mean-Field Decoupling} \label{maindecouple}
We define 8 bond operators as follows: 
\begin{subequations}
    \begin{align}
    \hat{\chi}_{ij} &= \sum_\alpha b^\dagger_{i\alpha} b_{j\alpha} \label{eq:bond1} \\
    \hat{E}^a_{ij} &= \sum_{\alpha\beta} b^\dagger_{i\alpha} [\sigma^a]_{\alpha\beta}b_{j\beta}\\
    \hat{\Delta}_{ij} &= \sum_{\alpha\beta} b_{i\alpha} [i\sigma^y]_{\alpha\beta}b_{j\beta}\\
    \hat{D}^a_{ij} &= \sum_{\alpha\beta} b_{i\alpha} [i\sigma^y\sigma^a]_{\alpha\beta} b_{j\beta} \label{eq:bond4},
    \end{align}    
\end{subequations}
where $a \in \{x,y,z\}$. The operators $\hat{\chi}, \hat{\vec{E}}$ are known as the singlet/triplet hopping, and $\hat{\Delta}, \hat{\vec{D}}$ are known as the singlet/triplet pairing. After replacing the pseudospins with Schwinger bosons, we write the Hamiltonian in the form \begin{align}
    H &= -\sum_{\la ij\ra} \boldsymbol{B}_{ij}^\dagger \xi_{ij} \boldsymbol{B}_{ij} \label{eq:hamb}
\end{align} where \begin{align}
    \boldsymbol{B}_{ij} &= (\hat{\chi}, \hat{E}^x, \hat{E}^y, \hat{E}^z, \hat{\Delta}, \hat{D}^x, \hat{D}^y, \hat{D}^z)_{ij}^\intercal
\end{align} and $\xi_{ij} \in \R^{8\times 8}$ is a real symmetric matrix to ensure $H$ is Hermitian. The decoupling matrix $\xi_{ij}$ is not determined uniquely, but we choose it to be positive-definite for reasons explained in appendices \ref{appdecouple} and \ref{appgeneralsc}. From Eq. \eqref{eq:hamb}, we are led to the mean-field decoupling \begin{align}
    H_\textrm{MF} &= -\sum_{\la ij\ra} (\boldsymbol{B}_{ij}^\dagger \xi_{ij} \boldsymbol{\Aa}_{ij} + \boldsymbol{\Aa}^\dagger_{ij} \xi_{ij} \boldsymbol{B}_{ij} - \boldsymbol{\Aa}^\dagger_{ij} \xi_{ij} \boldsymbol{\Aa}_{ij}) \label{eq:hambmf},
\end{align} 
where we introduce a vector $\boldsymbol{\Aa}_{ij} \in \C^8$ containing the mean-field amplitudes 
\begin{align}
    \boldsymbol{\Aa}_{ij} &= (\chi, E^x, E^y, E^z, \Delta,D^x, D^y, D^z)_{ij}^\intercal.
\end{align}
We stress that the components of $\boldsymbol{B}_{ij}$ are operators, but the components of $\boldsymbol{\Aa}_{ij}$ are variational parameters in the mean-field energy. The set of $\boldsymbol{\Aa}_{ij}$ will be referred to as an ansatz.

Regardless of the choice of decoupling, the mean-field Hamiltonian can be rewritten in the form \begin{align}
        H_{\textrm{MF}} &= - \sum_{\la ij\ra} \left( \boldsymbol{b}_i^{\dagger} \left[u^h_{ij}\right] \boldsymbol{b}_j + \boldsymbol{b}_i^{\intercal} \left[u^p_{ij}\right] \boldsymbol{b}_j\right) + \textrm{h.c.} \nonumber \\
        &+ \lambda \sum_i (n_i - \kappa) + \sum_{\la ij \ra} \boldsymbol{\Aa}^\dagger_{ij} \xi_{ij} \boldsymbol{\Aa}_{ij} \label{eq:hammfu}
\end{align} where $\boldsymbol{b}_i = (b_{i\up}, b_{i\dow})^\intercal$ is a vector of spinon operators, $u_{ij}^h, u_{ij}^p \in \C^{2\times 2}$, and we have introduced the site-independent Lagrange multiplier $\lambda$ to enforce the average occupation constraint \eqref{eq:occconstraint}. The matrices $u_{ij}^h$ and $u_{ij}^p$ are parameterized by complex numbers $a,b,c,d$ as follows: 
\begin{subequations}
\begin{align}
        u^h_{ij} &= a^h_{ij} \openone_{2\times2} + b^h_{ij} \sigma^x + c_{ij}^h \sigma^y + d_{ij}^h \sigma^z \label{eq:uhmatrix}\\
        u^p_{ij} &= i \sigma^y (a^p_{ij} \openone_{2\times2} + b^p_{ij} \sigma^x + c_{ij}^p \sigma^y + d_{ij}^p \sigma^z) \label{eq:upmatrix}
\end{align}     
\end{subequations}
In particular, the constants $a,b,c,d$ are linear combinations of the ansatz variables $\{\chi, E^x, E^y, E^z, \Delta, D^x, D^y, D^z\}$ with the coupling constants $J_{zz}, J_\pm, J_{\pm\pm},$ as coefficients.
\subsection{Projective Symmetry Group Classification}
In this section, we use the projective symmetry group (PSG) to classify the different mean-field ans\"atze. We are interested in ans\"atze which break no space group symmetries or time-reversal symmetry. However, the Schwinger boson representation contains a $U(1)$ gauge redundancy because the pseudospin operators \eqref{eq:schwingerboson} are invariant under the transformation \begin{equation}
    G: \boldsymbol{b}_{j} \mapsto e^{i\phi[j]} \boldsymbol{b}_{j} \label{eq:gauge}
\end{equation} Two ans\"atze which differ by a gauge transformation of the form \eqref{eq:gauge} are therefore physically equivalent~\cite{Wen2002, wen2004quantum, Wang2006}. The correct way to implement symmetry operations is via the gauge-enriched operations $\tilde{\Oo} = G_\Oo \circ \Oo$, where $\Oo$ is an element of the pyrochlore lattice space group \cite{Wen2002, Wang2006}. Explicitly, the action of $\tilde{\Oo}$ is \begin{equation}
    \tilde{\Oo}: \boldsymbol{b}_{j} \mapsto e^{i\phi_\Oo[\Oo(j)]} U_\Oo^\dagger \boldsymbol{b}_{\Oo(j)},
\end{equation} where $U_\Oo$ is a $2\times2$ unitary matrix representation of the space group operation. Similarly, the action of the gauge-enriched time reversal operation is \begin{align}
    \tilde{\mathcal{T}}: \boldsymbol{b}_{j} \mapsto e^{i\phi_{\mathcal{T}}[j]} \mathcal{K} U_\mathcal{T}^\dagger \boldsymbol{b}_j \label{eq:gtrs}
\end{align} with $\mathcal{K}$ being the complex conjugation operator and $U_\mathcal{T}$ is a $2\times2$ unitary matrix \cite{Liu2019}. Explicit expressions for $U_\Oo$ and $U_\mathcal{T}$ are derived in appendix \ref{apppseudotransform}; here we simply list the results \begin{subequations}
\begin{align}
    U_{T_1} = U_{T_2} =U_{T_3} &= \openone_{2\times2} \\ \label{eq:Tpseudotransform}
    U_{\overline{C}_6} &= \begin{pmatrix}
        \omega^* & 0 \\ 
        0 & \omega
    \end{pmatrix} \\
    U_{S} &= \begin{pmatrix}
        0 & \omega^* \\
        \omega & 0 
    \end{pmatrix} \\
    U_\mathcal{T} &= \begin{pmatrix}
        0 & 1 \\ 
        1 & 0
    \end{pmatrix} \label{eq:timepseudotransform}
\end{align}      
\end{subequations} where $\omega = e^{2\pi i /3}$. The group generated by the gauge-enriched operations $\tilde{\Oo}$ and $\tilde{\mathcal{T}}$ is known as the projective symmetry group (PSG) \cite{Wen2002, Wang2006, Liu2019}. Demanding that the mean-field Hamiltonian is invariant under the PSG elements puts constraints on the allowed mean-field ans\"atze. Ans\"atze that cannot be transformed into each other via a pure gauge transformation \eqref{eq:gauge} belong to different PSG classes, meaning they describe different QSLs \cite{Wen2002,Wang2006}. 

The subgroup of the PSG corresponding to pure gauge transformations associated with the identity operation $\Oo = \mathrm{id}$ is known as the invariant gauge group \cite{Wen2002,Wang2006}. The invariant gauge group corresponds to the emergent gauge structure of the resulting QSLs~\cite{Wen2002, wen2004quantum}. In this paper, we only consider the $\Z_2 = \{e^{i\pi n} | n \in \{0,1\}\}$ invariant gauge group, thus restricting our study to $\Z_2$ symmetric bosonic QSLs. 

Previous work on $\Z_2$ Schwinger boson mean-field theory has identified $16$ possible $\Z_2$ symmetric bosonic QSLs, with the phase factors $\phi_G$ given by 
\begin{subequations} \label{eq:psgsol}
    \begin{align}
        \phi_{T_1}(\rr_\mu) &= 0 \\
        \phi_{T_2}(\rr_\mu) &= n_1 \pi r_1 \\
        \phi_{T_3}(\rr_\mu) &= n_1 \pi (r_1+r_2) \\
        \phi_{\overline{C}_6}(\rr_\mu) &= \left[\frac{\nc}{2} + (n_1 + \nst) \delta_{\mu=1,2,3}\right]\pi \nonumber \\
        &+ n_1 \pi r_1 \delta_{\mu=2,3} + n_1\pi r_3 \delta_{\mu=2} \nonumber\\
        &+ n_1\pi r_1(r_2+r_3)\\
        \phi_S(\rr_\mu) &= \left[ (-1)^{\delta_{\mu=1,2,3}} \frac{n_1+\nst}{2} + \delta_{\mu=2} \ncs\right]\pi \nonumber \\
        &+ (n_1\delta_{\mu=1,2} + \nst) \pi r_1 \nonumber\\
        &+ (n_1 \delta_{\mu=2} + \nst) \pi r_2 + n_1\pi r_3 \delta_{\mu=1,2} \nonumber\\
        &+\frac{1}{2}n_1\pi(r_1+r_2)(r_1+r_2+1)\\
        \phi_{\mathcal{T}}(\rr_\mu) &= 0
    \end{align}     
\end{subequations}
where $n_1, n_{\overline{C}_6S}, n_{ST_1}, n_{\overline{C}_6}$ are all either $0$ or $1$ \cite{Liu2019}. Even though the PSG solution is the same as the effective spin-$1/2$~\cite{Liu2019} and dipolar-octupolar cases~\cite{Desrochers2022}, the structure of the allowed mean-field ans\"atze will be different because of the pseudospin transformation matrices $U_\Oo$ and $U_\mathcal{T}$. Following \cite{Liu2019}, we shift the first Brillouin zone $\qq \mapsto \qq - (\pi,\pi,\pi)$ for the PSG classes with $\nst = 1$, which is equivalent to a gauge transformation and does not affect physical quantities on the spin level. 

The PSG classes with $n_1 = 1$ have an enlarged unit cell because the translations $T_2$ and $T_3$ act projectively, which means that the translation-invariance of the ansatz is realized after projecting from the Schwinger boson Hilbert space down to the physical spin Hilbert space. We do not consider these classes in this work. There is no physical reason to exclude them, but only considering $n_1=0$ simplifies the calculations and the presentation of results. We therefore label the different $n_1=0$ PSG classes using the notation $(n_{\overline{C}_6S} n_{ST_1} n_{\overline{C}_6})$.  

\subsection{Generalized Self-Consistent Equations}
The self-consistent equations for our ansatz are determined by extremizing the mean-field energy $\la H_{MF} \ra$ with respect to the amplitudes $\Aa$ and the Lagrange multiplier $\lambda$. For the latter, we obtain the average occupation constraint \eqref{eq:occconstraint}; \begin{align}
    \frac{\di \la H_\textrm{MF}\ra}{\di \lambda} = 0 &\iff \la n_i \ra = \kappa.
\end{align} Assuming that each bond of the ansatz is independent and that $\xi$ is invertible, minimizing the mean-field energy with respect to each mean-field amplitude results in the self-consistent equations \begin{align}
    \la \boldsymbol{B}_{ij} \ra &= \boldsymbol{\Aa}_{ij} \label{eq:oldsc}
\end{align} for every bond $(i,j)$. The approach taken in previous work was to search for a subset of solutions to \eqref{eq:oldsc} that respected the PSG constraints \cite{Wang2006, Schneider2022, Desrochers2022}. 

However, our insistence that the ansatz respects the PSG constraints means this approach does not hold in all situations. When we impose the PSG constraints, the bonds are no longer independent. In fact, we only need to specify the MF parameters on one bond to fully determine their values on all other bonds. Thus, the derivation of \eqref{eq:oldsc} is modified, as described in more detail in appendix \ref{appgeneralsc}. We find a set of generalized self-consistent equations of the form \begin{align}
    \la \boldsymbol{B}_{i_0,j_0} \ra &= \mathcal{M} \boldsymbol{\Aa}_{i_0,j_0} \label{eq:generalgeneralsc}
\end{align} where the matrix $\mathcal{M} \in \R^{8\times 8}$ is determined by the decoupling matrix $\xi$ and the PSG. We note that this equation only holds for the reference bond $(i_0,j_0) = (\boldsymbol{0}_0,\boldsymbol{0}_1)$; the values of $\la \boldsymbol{B}_{ij}\ra$ and $\boldsymbol{\Aa}_{ij}$ for other bonds are fully determined by the MF values on the reference bond and the PSG. For simple choices of the decoupling, it can happen that $\mathcal{M} = \openone_{8\times8}$, recovering the usual result \eqref{eq:oldsc}. We find that it is not possible to choose such a decoupling for our Hamiltonian \eqref{eq:pseudospinham} (appendix \ref{appdecouple}), so the self-consistent equations always take a slightly more complicated form for the amplitudes $E^x$, $E^y$, $D^x$, and $D^y$. 

\section{Results}
\begin{figure*}[htbp]
\includegraphics[width=0.83\textwidth]{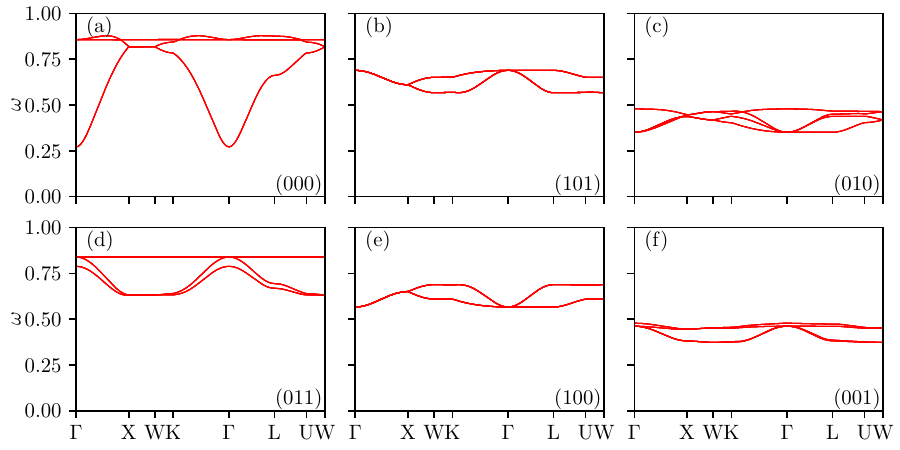}
\caption{\label{fig:disp} Spinon dispersion along a high symmetry path in the first Brillouin zone for phases (a) (000), (b) (101), (c) (010), (d) (011), (e) (100), and (f) (001). Every phase has a relatively flat dispersion with narrow bandwidth except for $(000)$, which is also the most susceptible to condensation.}  
\end{figure*}

\subsection{Phase Diagram}\label{results phase}
We compute the mean-field phase diagrams for $\kappa=0.1$ and $J_{zz}^2 + J_\pm^2 + J_{\pm\pm}^2 = 1$ with $J_\pm < 0$ and $J_{\pm\pm}>0$, which are presented in Fig. \ref{fig:fig 1} (c) and (d). We chose a small value of $\kappa$ to obtain all the competing QSL phases, since some QSLs may be unstable to magnetic ordering as $\kappa$ increases. We discuss this further in sections \ref{phase properties} and \ref{discussion}. We also chose $J_\pm < 0$ to explore the region of the parameter space with frustrated in-plane interactions, where QSLs are more likely to arise. Indeed, numerical studies have shown that the region $J_\pm > 0$ for $J_\pm \ge J_{zz}=0.05$ and $J_{zz}>0$ favours an in-plane pseudospin ferromagnetic order, while the region $J_\pm > 0$ for $J_{zz} < 0$ favours in-plane or Ising-type pseudospin ferromagnetic order \cite{banerjee2008unusual, Kato2015, huang2018dynamics, huang2020extended, Taillefumier2017, benton2018quantum, Schaden2025, Chung2024}. We parameterize the exchange couplings by 
\begin{equation}
    J_{zz} = \sin\theta\cos\phi, \ J_\pm = -\cos\theta, \ J_{\pm\pm} = \sin\theta\sin\phi
\end{equation} where $\theta \in (0,\frac{\pi}{2})$ and $\phi \in (0, \pi)$. As such, $\phi < \pi/2$ corresponds to $J_{zz}>0$ while $\phi> \pi/2$ corresponds to $J_{zz}<0$.

Since we have specialized to $n_1=0$, there are at most $8$ PSG classes to consider. For the range of parameters we studied, we found that certain PSG classes had identical mean-field ans\"atze in the regime of interest. Namely, the ans\"atze $(100)$ and $(110)$ were identical, as well $(101)$ and $(111)$, reducing the number of PSG classes to $6$. Strictly speaking, ans\"atze belonging to distinct PSG classes are still inequivalent, but we will treat them as the same since the resulting observables are the same. We expect this degeneracy to be lifted upon inclusion of other interactions or next-nearest neighbour terms in the mean-field ans\"atze.

We find that the lowest-energy state is often accompanied by a closely competing state, with a relative difference in mean-field energy on the order of $10^{-4}$ to $10^{-2}$. Thus, we show both the lowest- and second-lowest-energy states in Fig. \ref{fig:fig 1} (c) and (d), respectively. We find that $(000)$ is favoured in the antiferromagnetic (AFM) Ising ($J_{zz}=1$) and $J_{\pm\pm}=1$ limits, $(101)$ is favoured in the AFM XY limit ($J_\pm=-1$), and $(010)$ is favoured in the ferromagnetic (FM) Ising ($J_{zz}=-1$) limit. The transition from $(000)$ to $(101)$ occurs roughly when $\theta < \pi/4$, which is when $J_\pm$ becomes the dominant coupling. 

\begin{table}[hbtp]
\caption{\label{table:params}
Points in parameter space where observables (spinon dispersion, equal time structure factor, dynamic structure factor) are evaluated in this work. These points are marked by $\times$'s in Fig. \ref{fig:fig 1} (c) and (d).}
\begin{ruledtabular}
\centering
\begin{tabular}{c | c | c | c}
PSG Class & $J_{zz}$ & $J_\pm$ & $J_{\pm\pm}$ \\
\hline 
(000), (011)& 0.65 & -0.38 & 0.65\\
(101), (100)& 0.27 & -0.92 & 0.27 \\
(010), (001)& -0.90 & -0.38 & 0.22
\end{tabular}
\end{ruledtabular}
\end{table}

We point out the phase transition near the antiferromagnetic Heisenberg point $J_\pm/J_{zz} = -\frac{1}{2}$, $J_{\pm\pm}=0$ where $J_{zz}>0$, which is expected at the classical level \cite{Taillefumier2017, Rau2019}. Cluster-variational and pseudo-Majorana fermion functional renormalization group studies of the XXZ model on the pyrochlore lattice also support a phase transition to a magnetically disordered state there as well \cite{benton2018quantum,Schaden2025}, which supports our finding. Figure \ref{fig:fig 1} (d) shows that the phases $(011)$, $(100)$, and $(001)$ are the second lowest energy phases competing with $(000)$, $(101)$, and $(010)$ respectively. There is also a small region where $(101)$ has the second lowest energy compared to $(000)$. Excluding that small region, the pairs of lowest- and second-lowest-energy PSG classes have the same non-vanishing mean-field parameters $a^h, \ldots, d^h, a^p, \ldots, d^p$, defined in Eqs. (\ref{eq:uhmatrix}-\ref{eq:upmatrix}). However, they can still have distinct physical properties since they have different PSG parameters $\ncs, \nst, \nc$ which govern the spatial structure of the ans\"atze. Examples of this are shown in the next two sections.

\subsection{Distinguishing Properties of Different PSG Classes} \label{phase properties}

\begin{figure*}[hbtp]
\includegraphics[width=0.75\textwidth]{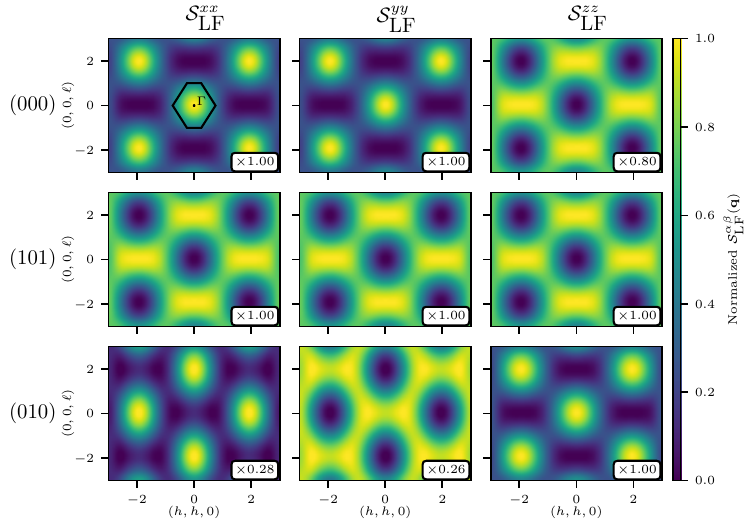}
\caption{\label{fig:local ssf} Local frame static spin structure factor components in the $[hhl]$ plane. Rows indicate the PSG class (000), (101), (010), and columns indicate the $xx$, $yy$, and $zz$ components of $\mathcal{S}^{\alpha\beta}_\textrm{LF}$ respectively. We set the minimum intensity of each panel to $0$ and normalized them independently for better visibility. The inset text shows the normalization factor for each panel relative to its row. For example, in the $(000)$ row, $\mathcal{S}_\textrm{LF}^{zz}$ has $0.8$ times the contrast as $\mathcal{S}_\textrm{LF}^{xx}$. Momenta are expressed in reciprocal lattice units, and the first Brillouin zone is indicated in black (top left panel).}  
\end{figure*}

Here we describe the different phases occurring in Fig. \ref{fig:fig 1} in more detail. For the rest of this work, all computations for a given PSG class are done using the self-consistent mean-field ans\"atze at the points marked by the $\times$'s in Fig. \ref{fig:fig 1} (c) and (d), and given in table \ref{table:params}.

The spinon dispersions of the six PSG classes are shown in Fig.~\ref{fig:disp}. As discussed in section \ref{results phase}, we chose $\kappa = 0.1$ so that all QSLs are gapped. However, we note that $(000)$ has a significantly larger bandwidth and smaller energy gap than the other classes, indicating its susceptibility to Schwinger boson condensation. In fact, for $\kappa \approx 0.14$, the class $(000)$ is condensed throughout the entire phase diagram. The minimum of its dispersion occurs at $\qq = 0$, consistent with the ordering wave vector expected from the classical phase diagram \cite{Rau2019,Chung2024,yan20017theory}. The smaller bandwidth and larger gap of the other PSG classes make them more resistant to condensation, making them more likely to persist at higher values of $\kappa$.

The different PSG classes are distinguishable through their local frame equal-time spin structure factors $\mathcal{S}^{\alpha\beta}_\textrm{LF}(\qq)$. These are defined by the spin-spin correlations 
\begin{align}
    \mathcal{S}^{\alpha\beta}_{\textrm{LF}}(\qq) &= \frac{1}{ N_\textrm{uc}}\sum_{\rr_\mu, \rr'_\nu} e^{i \qq \cdot(\rr_\mu-\rr'_\nu)} \la S_{\rr_\mu}^\alpha S_{\rr'_\nu}^\beta\ra \label{eq:localssf}
\end{align} and are shown in Fig. \ref{fig:local ssf}. The qualitative behaviour of $\mathcal{S}^{\alpha\beta}_\textrm{LF}(\qq)$ can be understood by looking at which mean-field amplitudes are nonzero after solving the generalized self-consistent equations \eqref{eq:generalgeneralsc}. The simplest are $(101)$ and $(100)$, which only have a nonzero singlet pairing $\Delta$ everywhere in the phase diagram. Since all triplet terms vanish, this gives them an emergent $SU(2)$ symmetry, so $\mathcal{S}^{xx}_{\textrm{LF}} = \mathcal{S}^{yy}_{\textrm{LF}} = \mathcal{S}^{zz}_{\textrm{LF}}$ for these two PSG classes. The intensity of $\mathcal{S}_\textrm{LF}$ has has maxima at nonzero $\qq$ ($[002]$ and symmetry-related points), indicating antiferromagnetic correlations. This is consistent with the fact that the antiferromagnetic $J_\pm$ is the dominant coupling when $(101)$ and $(100)$ have the lowest energy.

When $J_{zz}>0$, the classes $(000)$ and $(011)$ have nonzero singlet hopping $\chi$ and triplet pairing $D^z$. Since all the $x$ and $y$ triplet terms vanish, there is an emergent $U(1)$ symmetry that ensures $\mathcal{S}^{xx}_{\textrm{LF}} = \mathcal{S}^{yy}_{\textrm{LF}}$. The intensity of $\mathcal{S}^{zz}_\textrm{LF}$ has maxima at nonzero $\qq$ ($[002]$ and symmetry-related points), which is consistent with the antiferromagnetic $J_{zz}$ coupling. 

On the other hand, the self-consistent ans\"atze in the $J_{zz}<0$ region for $(000)$ and $(011)$ have nonzero $\chi, D^x, D^y$, which include $x$ and $y$ triplet mean-field amplitudes. The same is true for $(010)$ and $(001)$. As such, there is no emergent symmetry in these phases, and the spin-spin correlations are different for each component. In contrast to the previous two classes, the intensity of $\mathcal{S}^{zz}_\textrm{LF}$ has a maximum at $[000]$, which is consistent with the ferromagnetic $J_{zz}$ coupling. Figure \ref{fig:local ssf} shows that $\mathcal{S}^{zz}_\textrm{LF}$ is significantly stronger than the in-plane correlations for $(010)$ and $(001)$, but weaker than the in-plane correlations for $(000)$ and $(011)$. 

\subsection{Dynamic Spin Structure Factor}
The observable we are primarily interested in is the dynamic spin structure factor (DSSF) $\mathcal{S}$, which is directly related to the inelastic neutron scattering cross section \cite{Book2011, Boothroyd2020}. The quadrupolar $S^x$ and $S^y$ do not couple to the neutron's spin~\cite{chen2017dirac}, so neutron scattering only probes correlations between $S^z$. The resulting DSSF is
\begin{align}
    \mathcal{S}(\qq,\omega) &= \sum_{\mu\nu} \left(\hat{z}_\mu \cdot \hat{z}_\nu - \frac{(\hat{z}_\mu \cdot \qq)(\hat{z}_\nu \cdot \qq)}{|\qq|^2}\right)\mathcal{S}^{zz}_{\textrm{LF},\mu\nu}(\qq,\omega)
\end{align}
where the prefactor includes the transverse projector and the rotation from the pseudospin local frames to the global frame in Cartesian coordinates. The local frame DSSF is defined as
\begin{align}
    \mathcal{S}^{\alpha\beta}_{\textrm{LF},\mu\nu}(\qq,\omega) &= \frac{1}{2\pi N_\textrm{uc}} \int \dd t e^{i \qq \cdot(\rr_\mu-\rr'_\nu) + i \omega t} \la S_{\rr_\mu}^\alpha(t) S_{\rr'_\nu}^\beta(0)\ra. \label{eq:localdssf}
\end{align} 
Note that the local frame equal-time spin structure factor \eqref{eq:localssf} is obtained from Eq. \eqref{eq:localdssf} by integrating over $\omega$.

\begin{figure}
    \includegraphics[width=\linewidth]{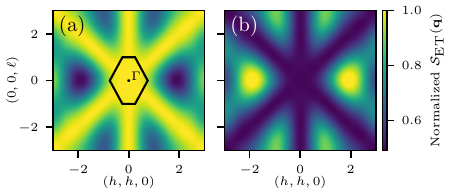}
    \caption{\label{fig:ssf} The equal-time neutron scattering amplitude $\mathcal{S}$ in the $[hhl]$ plane for (a) $(000), (011)$ when $J_{zz}>0$ and $(101), (100)$ everywhere, and (b) $(000), (011)$ when $J_{zz}<0$ and $(010), (001)$ everywhere. Momenta are expressed in reciprocal lattice units, and the first Brillouin zone is indicated in black in panel (a).} 
\end{figure}

We can gain some qualitative insight by first computing the equal-time structure factor (relevant for energy-integrated neutron scattering experiments) 
\begin{equation}
    \mathcal{S}_\textrm{ET}(\qq) = \int \mathcal{S}(\qq,\omega) \dd \omega.   
\end{equation}
We find a high-intensity ``rod" pattern (Fig. \ref{fig:ssf} (a)) for the PSG classes $(000)$ and $(011)$ only when $J_{zz}>0$. They also occur for the PSG classes $(101)$ and $(100)$ for both signs of $J_{zz}$. The high-intensity rods in the $[hhl]$ plane resemble the washed-out pinch point structure observed in experiments \cite{Kimura2013,Petit2016,Wen2017}. Such rods appear in single tetrahedron calculations of the equal-time structure factor \cite{Castelnovo2019}. On the other hand, we find low-intensity rods (Fig. \ref{fig:ssf} (b)) for $(000)$ and $(011)$ when $J_{zz}<0$, as well as for $(010)$ and $(001)$ regardless of the sign of $J_{zz}$. Since the high-intensity pattern better matches what is experimentally observed, we restrict our attention to the phases $(000), (011)$ (in the region $J_{zz}>0$), and $(101), (100)$ for the rest of this work. 

\begin{figure*}[hbtp]
\includegraphics[width=0.75\textwidth]{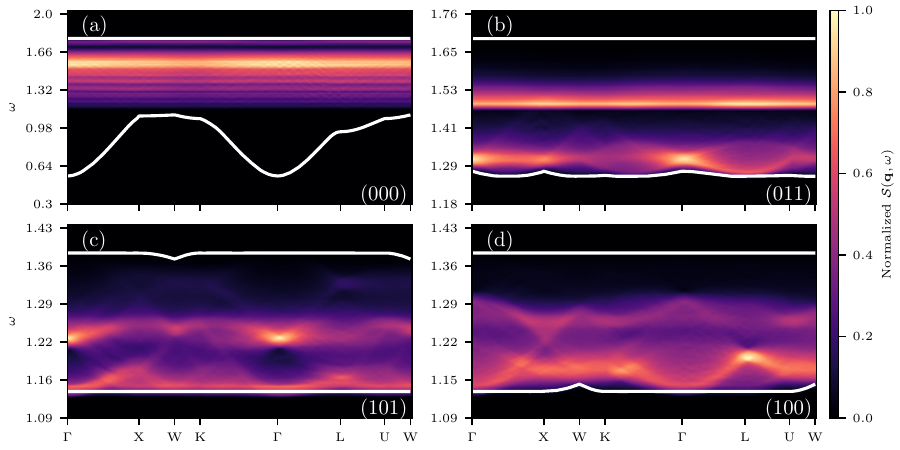}
\caption{\label{fig:dssf}Normalized dynamical spin structure factor (DSSF) along a high symmetry path in the first Brillouin zone for the lowest and second lowest energy $J_{zz}>0$ candidates: (a) (000), (b) (011), (c) (101), (d) (100). The lower and upper edges of the two-spinon continuum are indicated in white. Note that panels do not share the same energy axis.} 
\end{figure*}

We show the full dynamic spin structure factor and lower/upper edges of the two-spinon continuum along a high symmetry path in the first Brillouin zone in Fig. \ref{fig:dssf}. The two-spinon continuum at $\qq$ is defined as the energy interval
\begin{equation}
    \min_{\kk \in \textrm{BZ}} (\omega_\kk + \omega_{\qq-\kk}) \leq \omega \leq \max_{\kk \in \textrm{BZ}} (\omega_\kk + \omega_{\qq-\kk}),    
\end{equation} where $\omega_\kk$ is the spinon dispersion (see Fig. \ref{fig:disp}). Physically, a neutron that excites two spinons and is scattered with momentum transfer $\qq$ must transfer an energy $\omega$ that lies within the two-spinon continuum, at least at the mean-field level with non-interacting spinons. The DSSF is therefore $0$ outside the two-spinon continuum. The DSSFs of most candidates include a relatively flat feature where most of the spectral weight is concentrated, which is expected from the flatness of the spinon bands as shown in Fig. \ref{fig:disp}. These DSSFs are broadly consistent with the inelastic signal observed in experiments, but the resolution of currently available instruments may not be fine enough to show any of the detailed features in Fig. \ref{fig:dssf}. Interestingly, the DSSF for $(000)$ has a near-uniform $\qq$-dependence, which we attribute to the vanishing of the mean-field amplitudes $D^x$ and $D^y$ after reaching self-consistency, causing $c^p=0$ (as defined in Eq. \ref{eq:upmatrix}). In fact, $c^p \neq 0$ when $J_{zz}<0$ for $(000)$, and a nontrivial $\qq$-dependence in the DSSF is seen. 

\begin{figure*}[hbtp]
\includegraphics[width=0.83\textwidth]{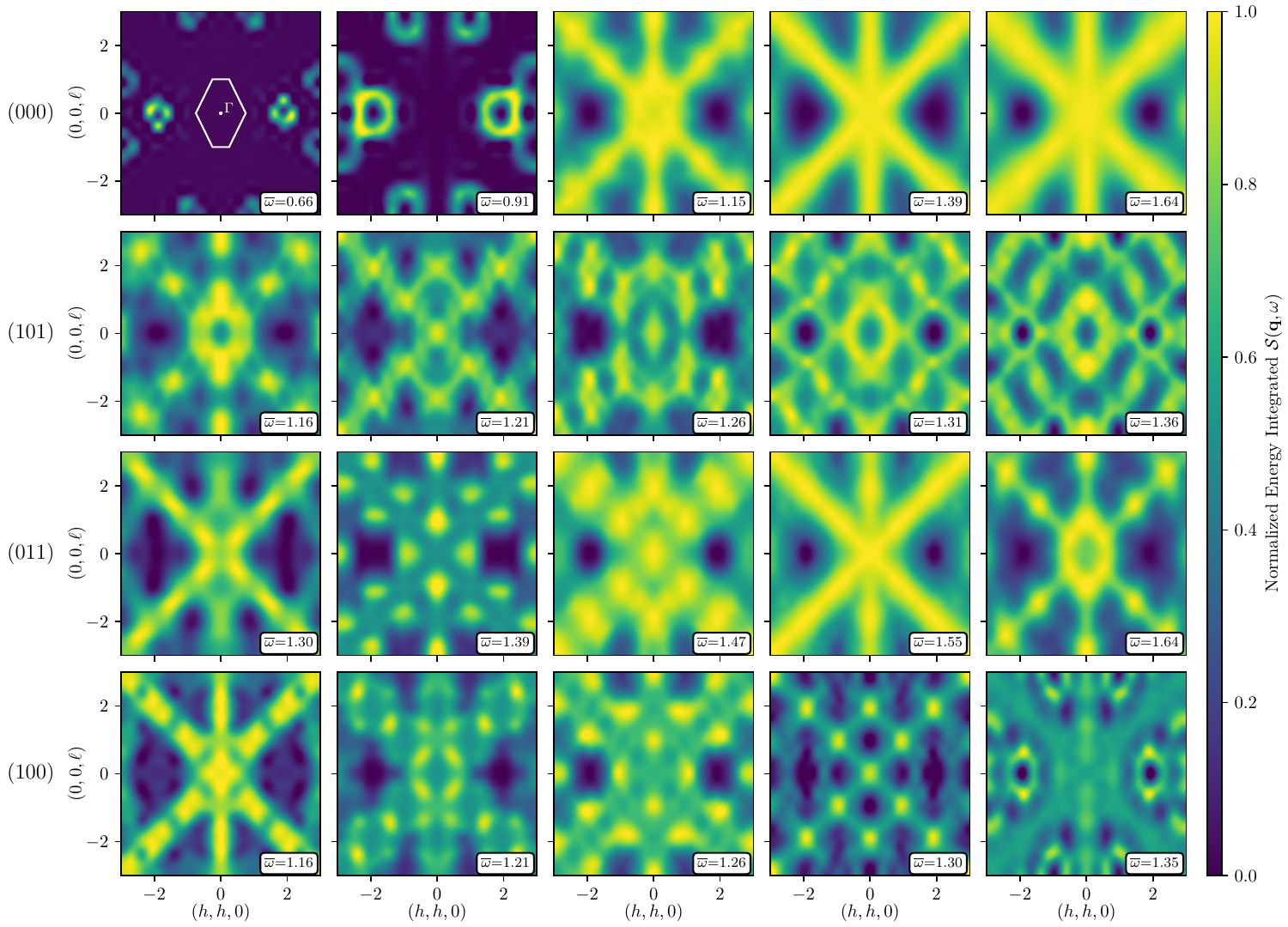}
\caption{\label{fig:all cuts}Energy-integrated DSSF in the $[hhl]$ plane for the lowest and second lowest energy candidates; $(000), (101), (011)$, and $(100)$ in the first, second, third and fourth rows, respectively. Each panel shows $\mathcal{S}(\qq,\omega)$ integrated over the energy interval $(\overline{\omega}-\delta\omega, \overline{\omega}+\delta\omega)$ for different $\overline{\omega}$ (inset text) but the same $\delta\omega$ in each row. Together, the $5$ intervals cover the entire two-spinon continuum. We set the minimum intensity of each panel to $0$ and normalized them independently for better visibility. Momenta are expressed in reciprocal lattice units, and the first Brillouin zone is shown in white (top left panel).} 
\end{figure*}

We also show energy cuts of $\mathcal{S}(\qq,\omega)$ integrated over energy windows $(\overline{\omega}-\delta\omega,\overline{\omega}+\delta\omega)$ in the $[hhl]$ plane in Fig. \ref{fig:all cuts}. Together, the windows cover the entire two-spinon continuum. 
Due to the flat features seen in the momentum-resolved DSSF (Fig. \ref{fig:dssf}), the energy window $\overline{\omega}$ centered around the flat feature will have the highest intensity. The cleanest realization of the rod structure occurs in the higher energy windows of $(000)$, but it can also be seen in the other candidates. Some energy-integrated windows display significant modulation of the rods, which is similar to what has been observed in $\prhfo$ \cite{Sibille2018}. 

\section{Discussion} \label{discussion}
We investigated possible QSL phases beyond the quantum spin ice regime for pyrochlore oxides with non-Kramers magnetic ions. Local moments are described by pseudospin degrees of freedom consisting of dipolar Ising and quadrupolar XY moments. We used a Schwinger boson mean-field theory and the corresponding projective symmetry group to classify $\Z_2$ quantum spin liquids on the pyrochlore lattice with non-Kramers ions. By constructing a mean-field phase diagram, we identified four candidate QSLs in the frustrated region of parameter space that show a high intensity rod structure in the $[hhl]$ plane, consistent with the inelastic signal in neutron scattering experiments \cite{Kimura2013,Petit2016,Sibille2018}. Certain energy-integrated windows of the dynamic spin structure factor show a modulated rod structure that is reminiscent of the features in neutron scattering experiments.

Using our PSG classification scheme, the lowest energy ans\"atze that are consistent with experiment are the states $(000)$ and $(101)$, but it is important to keep in mind that $(011)$ and $(100)$ are closely competing states in the relevant parameter space. These states can be distinguished by the intensity patterns in their dynamical spin structure factors.

In Schwinger boson mean-field theory, we use $\kappa$, the average boson occupation number per site, as a control parameter for quantum fluctuations. Smaller $\kappa$ means larger quantum fluctuations. In order to obtain all the competing QSL phases, we set $\kappa=0.1$ in the mean-field phase diagram, which is much smaller than the physical value of $\kappa = 1$. Upon increasing $\kappa$, we find that the phase $(000)$ is much more susceptible to Schwinger boson condensation than the other three phases because of its large spinon bandwidth and small excitation gap at $\qq=0$. This suggests that the region covered by $(000)$ becomes magnetically ordered (with $\qq=0$ ordering wave vector) for realistic values of $\kappa$, unlike $(101)$ and $(100)$ which are more robust due to their narrow and relatively flat spinon bands, and large excitation gap. In this sense, $(101)$ and $(100)$ are stronger $\Z_2$ QSL candidates. 

Our phase diagram shows that the $\Z_2$ quantum spin liquids $(101)$ and $(100)$ are energetically favoured when $J_\pm$ is the dominant coupling. Interestingly, numerical studies of the XXZ model \eqref{eq:xxz_model} using cluster-variational and pseudo-Majorana fermion functional renormalization group calculations also find a transition to a disordered state past the antiferromagnetic Heisenberg point $J_\pm/J_{zz} < - \frac{1}{2}$ where $J_{zz}>0$ \cite{benton2018quantum, Schaden2025}. This supports the possibility that a quantum spin liquid could occur when $|J_\pm|$ becomes comparable to or larger than $|J_{zz}|$. Thus, it is possible that $(101)$ and $(100)$ are among several competing QSL phases in this region of parameter space. 

The non-Kramers doublet can be split by crystalline disorder or local strain because its degeneracy is protected only by crystalline symmetries, not by time-reversal symmetry. Previous work indicates that this plays an important role in $\przo$ physics and could explain the inelastic neutron scattering signal \cite{Martin2017, Wen2017, Savary2017, Benton2018}. It will be important to precisely determine the exchange couplings $J_{zz}, J_\pm, J_{\pm\pm}$ to reliably assess the effects of disorder. At the moment, these parameters are mostly unknown. In the context of Schwinger boson theory, future work could investigate the effects of disorder on candidate QSL phases. On the other hand, our analysis may be more directly applicable to $\prhfo$, which is claimed to have minimal disorder \cite{Sibille2018}. 

\begin{acknowledgments}
This work was supported by the Natural Science and Engineering Council of Canada (NSERC) Discovery Grant No. RGPIN-2023-03296 and the Center for Quantum Materials at the University of Toronto. Computations were performed on the Cedar cluster hosted by the Digital Research Alliance of Canada. T.A. is supported by the Canada Graduate Scholarship (CGS-M). F.D. is further supported by the Vanier Canada Graduate Scholarship (CGV-186886).
\end{acknowledgments}

\bibliography{non_kramers_schwinger_boson_bib}

\appendix

\section{Microscopic Model} \label{appmicroscopic}
Starting from Hund's rules, an $f^2$ state has total spin $S=1$ and orbital angular momentum $L=5$, leading to a total angular momentum of $J=4$. Decomposing the 9-dimensional representation $\Gamma_{J=4}$ of $SO(3)$ in terms of irreducible representations of the double group of $D_{3d}$, we find \begin{equation}
    \Gamma_{J=4} = 2A_{1g} \oplus A_{2g} \oplus 3 E_g
\end{equation} in terms of standard Mulliken notation \cite{Bradley2009}. The two-dimensional irreducible representation $E_g$ forms the non-Kramers doublet, with an explicit construction given in Eq. \eqref{eq:doublet} \cite{Rau2019}. 

The local basis vectors for the four sublattices of the pyrochlore lattice are shown in table \ref{table:localframes}. The bond-dependent phase factor $\g_{ij} = \g_{\rr_\mu, \rr'_\nu}$ depends only on the sublattice indices $(\mu,\nu)$, and is defined by 
\begin{align}
    \gamma_{\mu\nu} &= \begin{pmatrix}
        0 & 1 & \omega & \omega^2 \\
        1 & 0 & \omega^2 & \omega \\
        \omega & \omega^2 & 0 & 1 \\
        \omega^2 & \omega & 1 & 0
    \end{pmatrix}
\end{align} where $\omega = e^{2\pi i /3}$ \cite{Onoda2011}.
\begin{table}[hbtp]
\caption{\label{table:localframes}%
Local sublattice basis vectors expressed in global Cartesian coordinates.}
\begin{ruledtabular}
\begin{tabular}{c c c c c}
$\mu$ & 0 & 1 & 2 & 3 \\
\hline 
$\hat{z}_\mu$ & $\frac{1}{\sqrt{3}}(1,1,1)$ & $\frac{1}{\sqrt{3}}(1,-1,-1)$ & $\frac{1}{\sqrt{3}}(-1,1,-1)$ & $\frac{1}{\sqrt{3}}(-1,-1,1)$\\
$\hat{y}_\mu$ & $\frac{1}{\sqrt{2}}(0,-1,1)$ & $\frac{1}{\sqrt{2}}(0,1,-1)$ & $\frac{1}{\sqrt{2}}(0,-1,-1)$ & $\frac{1}{\sqrt{2}}(0,1,1)$\\
$\hat{x}_\mu$ & $\frac{1}{\sqrt{6}}(-2,1,1)$ & -$\frac{1}{\sqrt{6}}(2,1,1)$ & $\frac{1}{\sqrt{6}}(2,1,-1)$ & $\frac{1}{\sqrt{6}}(2,-1,1)$\\
\end{tabular}
\end{ruledtabular}
\end{table}
\section{Pseudospin Transformations} \label{apppseudotransform}
We show how to derive the matrices $U_\Oo$ for the space group generators and $U_\mathcal{T}$ for time-reversal. The transformation of the pseudospin under a space group generator can be deduced by first performing the active transformation on the pseudospin and then changing the local coordinates due to sublattice mixing. The representations of the generators $\overline{C}_6$ and $S$ acting on a pseudospin state $|\pm\ra_\mu$ on sublattice $\mu$ are
\begin{subequations}
    \begin{align}
        R_{\overline{C}_6, \mu} &= \exp\left(-i \frac{2\pi}{3} \boldsymbol{\hat{n}}_{\overline{C}_6,\mu} \cdot \boldsymbol{J}\right) \\
        R_{S, \mu} &= \exp\left(-i \pi  \boldsymbol{\hat{n}}_{S,\mu} \cdot \boldsymbol{J}\right) 
    \end{align}
\end{subequations}
where $\boldsymbol{\hat{n}}_{\Oo,\mu}$ is the rotation axis of the generator $\Oo$ written in the local frame $\mu$ and $\boldsymbol{J}$ is the angular momentum operator. For the ground state \eqref{eq:doublet}, we have $J=4$, but we note it is equivalent and simpler computationally to take the ground state doublet as $|\pm\ra = |m=\pm1\ra$ with $J=1$ \cite{Rau2019}. In global Cartesian coordinates, 
\begin{equation}
    \boldsymbol{\hat{n}}_{\overline{C}_6} = \frac{1}{\sqrt{3}}(1,1,1)_\textrm{global}, \quad \boldsymbol{\hat{n}}_{S} = \frac{1}{\sqrt{2}}(1,1,0)_\textrm{global}.
\end{equation}
Since the local frames differ on each sublattice, we define $SO(3)$ matrices $O_{\mu\to\nu}$ which map the local frame $\mu$ to the local frame $\nu$ by 
\begin{align}
    O_{\mu\to\nu} &= W_\nu W_\mu^\intercal
\end{align} 
where the columns of $W_\mu$ are the local frame basis vectors $(\hat{x}_\mu,\hat{y}_\mu,\hat{z}_\mu)$. We call the unitary matrices representing these rotations $R_{\mu\to\nu}$. Putting it together, 
\begin{align}
    U_{\Oo,\mu} &= \mathcal{P} R_{\mu\to\Oo(\mu)} R_{\Oo,\mu} \mathcal{P} \label{eq:pseudotransformmatrix}
\end{align} 
where $\mathcal{P}$ is the projection to the ground state doublet. Evaluating \eqref{eq:pseudotransformmatrix} explicitly, the matrices $U_{\Oo,\mu}$ are independent of the sublattice index, and are given in Eqs. (\ref{eq:Tpseudotransform}-\ref{eq:timepseudotransform}). 

For the time reversal operator $\mathcal{T} = \exp\left(i\pi J^y\right) \mathcal{K}$, the matrix $U_\mathcal{T}$ evaluates explicitly to
\begin{subequations}
\begin{align}
    U_\mathcal{T} &= \mathcal{P} \exp\left(i\pi J^y\right) \mathcal{P} \\
    &= \begin{pmatrix}
        0 & 1 \\ 
        1 & 0
    \end{pmatrix}.
\end{align}    
\end{subequations}

\begin{table*}
\caption{\label{table:bond_constraints_full}Relation between the mean-field parameters on the different bonds in the pyrochlore unit cell. The solution of the $\Z_2$ PSG is parameterized by $n_1,n_C, n_{CS}, n_{ST} \in \{0,1\}$, although we set $n_1=0$ in this work. Explicit expressions for the matrices $\mathcal{R}^\g_{ij}$ can be derived as follows: $\mathcal{R}_{ij}^\g$ is the $4\times 4$ matrix corresponding to the linear transformation of $(a,b,c,d)$ in the row $(i \to j)$, multiplied by the adjacent $\Z_2$ phase. For example, $\mathcal{R}_{1_+, 2_+}^h = \begin{pmatrix} 
    1 & & & \\
    & -\half & \sqth & \\
    & \sqth & \half & \\
    & & & 1
\end{pmatrix}(-1)^{n_1+n_{ST}}$. In this table only, we use that notation $\mu_\eta$ to specify sites within the pyrochlore unit cell at $\ze$, where $+$ (-) denotes a down- (up-) pointing tetrahedron; see Fig. \ref{fig:fig 1} (a). The mapping back to sublattice indexed pyrochlore coordinates is $\mu_\eta \mapsto \ze_\mu$ if $\eta = +$ and $\mu_\eta \mapsto (-\ee_\mu)_\mu$ if $\eta = -$.}
    \begin{ruledtabular}
    \begin{tabular}{|c|c|c|c|c|}
        Bond & Hopping  & $\Z_2$ phase & Pairing & $\Z_2$ phase\\
        \hline
        $0_+\to1_+$ & $(a^h,b^h,c^h,d^h) $ & $1$ & $(a^p,b^p,c^p,d^p)$ & $1$ \\[1ex]
        \hline
        $0_+\to2_+$ & $(a^h,-\half b^h + \sqth c^h,-\sqth b^h -\half c^h,d^h)$ & $1$ & $(a^p,-\half b^p + \sqth c^p, -\sqth b^p - \half c^p,d^p)$ & $1$ \\[1ex]
        \hline
        $0_+\to3_+$ & $(a^h,-\half b^h - \sqth c^h,\sqth b^h -\half c^h,d^h)$ & $1$ & $(a^p,-\half b^p - \sqth c^p, \sqth b^p - \half c^p,d^p)$ & $1$ \\[1ex]
        \hline
        $1_+\to2_+$ & $(a^h,-\half b^h + \sqth c^h, \sqth b^h + \half c^h,-d^h)$ & $(-1)^{n_1 + n_{ST}}$ & $(-a^p,\half b^p - \sqth c^p, -\sqth b^p - \half c^p,d^p)$ & $(-1)^{n_C}$ \\[1ex]
        \hline
        $1_+\to3_+$ & $(a^h,-\half b^h - \sqth c^h, -\sqth b^h + \half c^h,-d^h)$ & $(-1)^{n_1+ n_{CS} + n_{ST}}$ & $(-a^p,\half b^p + \sqth c^p, \sqth b^p - \half c^p,d^p)$ & $(-1)^{n_C+n_{CS}}$ \\[1ex]
        \hline
        $2_+\to3_+$ & $(a^h,b^h,-c^h,-d^h) $ & $(-1)^{n_1 + n_{ST}}$ & $(-a^p,-b^p ,c^p,d^p)$ & $(-1)^{n_C}$ \\[1ex]
        \hline 
        $0_+\to1_-$ & $(a^h,b^h,c^h,d^h) $ & $(-1)^{n_1 + n_{ST}}$ & $(a^p,b^p,c^p,d^p)$ & $(-1)^{n_1+ n_{CS} + n_{ST}}$ \\[1ex]
        \hline
        $0_+\to2_-$ & $(a^h,-\half b^h + \sqth c^h,-\sqth b^h -\half c^h,d^h)$ & $(-1)^{n_1 + n_{ST}}$ & $(a^p,-\half b^p + \sqth c^p, -\sqth b^p - \half c^p,d^p)$ & $(-1)^{n_1+ n_{CS} + n_{ST}}$ \\[1ex]
        \hline
        $0_+\to3_-$ & $(a^h,-\half b^h - \sqth c^h,\sqth b^h -\half c^h,d^h)$ & $(-1)^{n_1 + n_{ST}}$ & $(a^p,-\half b^p - \sqth c^p, \sqth b^p - \half c^p,d^p)$ & $(-1)^{n_1+ n_{CS} + n_{ST}}$ \\[1ex]
        \hline
        $1_-\to2_-$ & $(a^h,-\half b^h + \sqth c^h, \sqth b^h + \half c^h,-d^h)$ & $(-1)^{n_1 + n_{ST}}$ & $(-a^p,\half b^p - \sqth c^p, -\sqth b^p - \half c^p,d^p)$ & $1$ \\[1ex]
        \hline
        $1_-\to3_-$ & $(a^h,-\half b^h - \sqth c^h, -\sqth b^h + \half c^h,-d^h)$ & $(-1)^{n_1+ n_{CS} + n_{ST}}$ & $(-a^p,\half b^p + \sqth c^p, \sqth b^p - \half c^p,d^p)$ & $(-1)^{n_C+n_{CS}}$ \\[1ex]
        \hline
        $2_-\to3_-$ & $(a^h,b^h,-c^h,-d^h) $ & $(-1)^{n_1 + n_{ST}}$ & $(-a^p,-b^p ,c^p,d^p)$ & $1$ \\[1ex]
    \end{tabular}
    \end{ruledtabular}
\end{table*}

\section{Constraints on the Mean-Field Amplitudes} \label{psg constraints}
On the pyrochlore lattice, all nearest-neighbour bonds are symmetry-related. In this work, we will always take $(i_0,j_0)=(\boldsymbol{0}_0,\boldsymbol{0}_1)$ as the reference bond and denote it by $(0,1)$ for brevity. This lets us express all $u^\g_{ij}$ in terms of $u^\g_{01}$, corresponding to the reference bond. 

We demand that the gauge-enriched symmetry operations $\tilde{\Oo} = \G_\Oo \circ \Oo$ be symmetries of $H_{\textrm{MF}}$; 
\begin{subequations} 
    \begin{align}
        u^h_{\Oo i, \Oo j} &= U_\Oo u^h_{ij} U_\Oo^\dagger e^{i(-\phi_\Oo[\Oo i] + \phi_\Oo[\Oo j])} \label{eq:uhtransform} \\
        u^p_{\Oo i, \Oo j} &= U_\Oo^* u^p_{ij} U_\Oo^\dagger e^{i(\phi_\Oo[\Oo i] + \phi_\Oo[\Oo j])} \label{eq:uptransform}
    \end{align}     
\end{subequations}
where $\phi_\Oo$ are the phase factors from the PSG solution (\ref{eq:psgsol}). The phases $\phi_\Oo$ do not depend on the choice of ground state doublet, but the transformation matrices $U_\Oo$ do. Defining the vectors \begin{align}
    t^\g_{ij} &= \begin{pmatrix}
        a^\g_{ij} & b^\g_{ij} & c^\g_{ij} & d^\g_{ij}
    \end{pmatrix}^\intercal 
\end{align} where $\g \in \{h,p\}$ indicates hopping or pairing, the PSG constraints can be rewritten as \begin{align}
    t^\g_{\Oo i,\Oo j} &= \mathcal{R}^\g_\Oo t^\g_{ij} \label{eq:abcdtransform}
\end{align} where the $O(4)$ matrices $\mathcal{R}_\Oo^\g$ are determined by $U_\Oo$ and the phase factors in Eqs. (\ref{eq:uhtransform}-\ref{eq:uptransform}). This shows that it is enough to specify $u^\g_{ij}$ on a single bond $(i_0,j_0)$ to determine all the remaining $u^\g_{ij}$.Finally, since we have specialized to $\Z_2$ spin liquids, the phases are either $1$ or $-1$, so $(\mathcal{R}^\g_\Oo)^* = \mathcal{R}^\g_\Oo$. 

Defining the vector $\boldsymbol{t}_{ij} = (t^h_{ij}, t^p_{ij}) \in \C^8$, the transformation law \eqref{eq:abcdtransform} becomes \begin{align}
    \boldsymbol{t}_{ij} &= R_{ij} \boldsymbol{t}_{01}
\end{align} 
where the block-diagonal matrix $R_{ij} \in \R^{8\times 8}$ is
\begin{align}
    R_{ij} &= \begin{pmatrix}
        \mathcal{R}^h_{ij}  & \mathbf{0}_{4\times 4} \\
        \mathbf{0}_{4\times 4} & \mathcal{R}^p_{ij}
    \end{pmatrix}.
\end{align} 

Demanding that the gauge-enriched time reversal operation \eqref{eq:gtrs} is a symmetry of the mean-field Hamiltonian leads to the conditions 
\begin{subequations}
    \begin{align}
    u^p_{ij} &= U_\T^* \left[u^p_{ij}\right]^* U_\T^\dagger \\
    u^h_{ij} &= U_\T \left[u^h_{ij}\right]^* U_\T^\dagger. 
    \end{align} 
\end{subequations}
For the non-Kramers doublet, we have $U_\T = \sigma^x$, resulting in the constraints
\begin{subequations}
    \begin{align}
        (a^h, b^h, c^h, d^h) &= ((a^h)^*, (b^h)^*, (c^h)^*, -(d^h)^*) \label{eq:trsh}\\
        (a^p, b^p, c^p, d^p) &= (-(a^p)^*, - (b^p)^*, -(c^p)^*, (d^p)^*), \label{eq:trsp}
    \end{align}        
\end{subequations}
which hold for every bond $(i,j)$. Thus, $a^h, b^h, c^h, d^p$ are real and $a^p, b^p, c^p, d^h$ are imaginary.

The last set of constraints arises because a bond can be mapped to itself in several inequivalent ways \cite{Desrochers2022}. First, we note that the hopping and pairing mean-field amplitudes satisfy the following symmetry properties
\begin{subequations}
        \begin{align}
        \chi_{ji} &= \chi^*_{ij} \label{eq:switch1}\\
        E^a_{ji} &= (E^a_{ij})^* \\
        \Delta_{ji} &= - \Delta_{ij}\\
        D^a_{ji} &= D^a_{ij}, \label{eq:switch4}
    \end{align}     
\end{subequations} which follows from the definitions (\ref{eq:bond1}-\ref{eq:bond4}). A group theory analysis \cite{Desrochers2022} shows that
\begin{subequations}
    \begin{align}
        g_1 &= \mathrm{id} \\
        g_2 &= S \overline{C}_6 S \overline{C}_6^{-1} S \overline{C}_6^{-1} \\ 
        g_3 &= \overline{C}_6 S^{-1} \overline{C}_6^2 \\ 
        g_4 &= \overline{C}_6 S^{-1} \overline{C}_6^2 S \overline{C}_6 S \overline{C}_6^{-1} S \overline{C}_6^{-1}    
    \end{align}    
\end{subequations} are the four inequivalent ways that the undirected bond $(\boldsymbol{0}_0, \boldsymbol{0}_1)$ can be mapped to itself, which we will call stabilizers. Note that the stabilizers $g_3$ and $g_4$ reverse the orientation of the bond.   
    
Applying Eqs. (\ref{eq:uhtransform}-\ref{eq:uptransform}) with the stabilizers above and the properties (\ref{eq:switch1}-\ref{eq:switch4}), we find
\begin{subequations}
    \begin{align}
        (a^h,b^h,c^h, d^h) &= (-1)^{n_{CS}} \times (a^h, b^h, -c^h, -d^h) \\
        &= (a^h, b^h, -c^h, d^h) \\
        &= (-1)^{n_{CS}} \times (a^h, b^h, c^h, -d^h)
    \end{align}     
\end{subequations}
This implies that $c^h = 0$ regardless of PSG class. We see that if $n_{CS}=1$, all pairing terms vanish except $a^p \in i\R$, and all hopping terms vanish except $d^h \in i\R$. 
 
 For the case $n_{CS} = 0$, the remaining constraints are 
\begin{subequations}
   \begin{align}
        (a^p,b^p,c^p,d^p) &= (-1)^{n_1 + n_C + n_{ST}} (-a^p,-b^p,c^p,d^p) \\
        &= (-1)^{n_1 + n_C + n_{ST}} (a^p,-b^p,c^p,d^p),
   \end{align}     
\end{subequations}
which are equivalent to 
\begin{align}
   \begin{cases}
        a^p = b^p = 0 & \textrm{if } n_1+n_C+n_{ST} \equiv 0 \pmod{2}\\
        a^p = c^p = d^p = 0 & \textrm{if } n_1+n_C+n_{ST} \equiv 1 \pmod{2} 
   \end{cases}
\end{align}     
and $a^h, b^h \in \R$ are the nonzero hopping terms.

\section{Mean-Field Decoupling Scheme} \label{appdecouple}
After replacing the pseudospin operators with Schwinger bosons, the terms in the Hamiltonian \eqref{eq:pseudospinham} can be written in terms of bond operators and number operators $n_i = \sum_\alpha b^\dagger_{i\alpha} b_{i\alpha}$ as in Eqs. (\ref{eq:szdecouple}-\ref{eq:spmpmdecouple}). Every term contains the expression
\begin{align}
    F_{ij} &= A \chi^\dagger_{ij} \chi_{ij} + B \Delta^\dagger_{ij} \Delta_{ij} + C^x (E^x_{ij})^\dagger E^x_{ij} \nonumber \\
    &+ C^y (E^y_{ij})^\dagger E^y_{ij} + C^z (E^z_{ij})^\dagger E^z_{ij}, 
\end{align}
where $A,B, C^x,C^y,C^z,N$ are any real constants. We have reused these variables in Eqs. (\ref{eq:szdecouple}-\ref{eq:spmpmdecouple}) to reduce clutter, but it is understood that $A, B, C^x, C^y, C^z, N$ in different equations can be different. We drop the terms containing $n_j$ and $n_in_j$ in this work because they are constant at the mean-field level. 

We then choose the values of $A, B, C^x, C^y, C^z, N$  so that the decoupling matrix $\xi_{ij}$ is always positive-definite, meaning these values depend on the signs of $J_{zz}$ and $J_\pm$. This is done so that the mean-field Hamiltonian is bounded below in the amplitudes $\chi, E^x, \ldots, D^z$, and also to ensure $\xi_{ij}$ is always invertible. The boundedness criterion is not strictly necessary, but it allows one to use numerical methods such as simulated annealing to solve the self-consistent equations. However, it is important that $\xi_{ij}$ is invertible for the generalized self-consistent equations (appendix \ref{appgeneralsc}) to make sense.

The decoupling is summarized in table \ref{table:fulldecoupling}. We see that it is not possible to choose a decoupling without ``off-diagonal terms" (i.e. terms of the form $(E^x)^\dagger E^y$) in Eq. \eqref{eq:spmpmdecouple}, which is relevant for which amplitudes satisfy generalized self-consistent equations (appendix \ref{appgeneralsc}). We also note that the choice of decoupling is not unique.
\begin{widetext}
\begin{subequations}
    \begin{align}
        4 S^z_i S^z_j &=F_{ij} + (1-A+B+C^x) (D^x_{ij})^\dagger D^x_{ij} + (1-A+B+C^y) (D^y_{ij})^\dagger D^y_{ij} \nonumber\\
        &+ (-A+B+C^z) (D^z_{ij})^\dagger D^z_{ij} + (-A-C^x-C^y-C^z) n_j \nonumber\\
        &+ (-1+A-2B-C^x-C^y-C^z) n_i n_j \label{eq:szdecouple} \\
        2(S^+_i S^-_j + S^-_i S^+_j)  &=F_{ij} + (1-A+B+C^x) (D^x_{ij})^\dagger D^x_{ij} + (1-A+B+C^y) (D^y_{ij})^\dagger D^y_{ij}\nonumber\\
        &+ (2-A+B+C^z) (D^z_{ij})^\dagger D^z_{ij} + (-A-C^x-C^y-C^z) n_j \nonumber\\
        &+ (-2+A-2B-C^x-C^y-C^z) n_i n_j \label{eq:spmdecouple}\\
        \g_{ij} S^+_i S^+_j + \g_{ij}^* S^-_i S^-_j &= F_{ij} + (-A+B+C^x - \frac{1}{2}(\textrm{Re}\g_{ij} + \textrm{Im} \g_{ij})) (D^x_{ij})^\dagger D^x_{ij} \nonumber\\
        &+ (-A + B + C^y + \frac{1}{2}(\textrm{Re}\g_{ij} - \textrm{Im} \g_{ij}))(D^y_{ij})^\dagger D^y_{ij} + (-A+B+C^z) (D^z_{ij})^\dagger D^z_{ij} \nonumber\\
        &+ \frac{1}{2}(-A-C^x-C^y-C^z-N+\textrm{Im} \g_{ij}) (D^x_{ij}+D^y_{ij})^\dagger (D^x_{ij} + D^y_{ij}) \nonumber\\
        &+ \frac{1}{2}(-A-C^x-C^y-C^z-N) (E^x_{ij}+E^y_{ij})^\dagger (E^x_{ij} + E^y_{ij}) + N n_j + (2A-2B+N) n_i n_j, \label{eq:spmpmdecouple}
    \end{align} 
\end{subequations}
\end{widetext}

\begin{table*}
\caption{\label{table:fulldecoupling}%
The nonzero entries of the $L^\g_{ij}$ matrices making up the decoupling matrix $\xi_{ij}$ (Eq. \eqref{eq:ximatrix}) depending on the sign of the exchange couplings. We are only considering the case $J_\pm < 0$. We take $J_{\pm\pm}>0$ without loss of generality.}
\begin{ruledtabular}
\begin{tabular}{|c|c|}
$J_{zz} > 0$ & $J_{zz} <0$\\
\hline 
$\begin{aligned}
    \relax[L^h_{ij}]_{1,1} &= |J_\pm|\\
    [L^h_{ij}]_{2,2} =[L^h_{ij}]_{3,3}&= \tfrac{1}{4}|J_{zz}| + |J_\pm|+\left(1+\tfrac{1}{4}\textrm{Im} \g_{ij}\right)|J_{\pm\pm}| \\
    [L^h_{ij}]_{2,3} = [L^h_{ij}]_{3,2} &= \left(1+\tfrac{1}{4}\textrm{Im} \g_{ij}\right)|J_{\pm\pm}| \\
    [L^h_{ij}]_{4,4} &= \tfrac{1}{4}|J_{zz}| + \tfrac{3}{2}|J_\pm| + \tfrac{1}{2}|J_{\pm\pm}| \\
    [L^p_{ij}]_{1,1} &= |J_\pm|+ |J_{\pm\pm}|\\
    [L^p_{ij}]_{2,2} =[L^p_{ij}]_{3,3}&= \tfrac{1}{2}|J_\pm|+\left(2+ \tfrac{1}{2}\textrm{Re}\g_{ij}+\tfrac{1}{2}\textrm{Im} \g_{ij}\right)|J_{\pm\pm}| \\
    [L^p_{ij}]_{2,3} = [L^p_{ij}]_{3,2} &= \left(1+\tfrac{1}{4}\textrm{Im} \g_{ij}\right)|J_{\pm\pm}| \\
    [L^p_{ij}]_{4,4} &= \tfrac{1}{4}|J_{zz}| + \tfrac{1}{2}|J_\pm| + \tfrac{3}{2}|J_{\pm\pm}|
\end{aligned}$ &
$\begin{aligned}
    \relax[L^h_{ij}]_{1,1} &= \tfrac{1}{4}|J_{zz}|\\
    [L^h_{ij}]_{2,2} =[L^h_{ij}]_{3,3} &= \tfrac{1}{4}|J_{zz}| + \left(1+\tfrac{1}{4}\textrm{Im} \g_{ij}\right)|J_{\pm\pm}| \\
    [L^h_{ij}]_{2,3} = [L^h_{ij}]_{3,2} &= \left(1+\tfrac{1}{4}\textrm{Im} \g_{ij}\right)|J_{\pm\pm}| \\
    [L^h_{ij}]_{4,4} &= \tfrac{1}{4}|J_{zz}| + \tfrac{1}{2}|J_\pm| + \tfrac{1}{2}|J_{\pm\pm}| \\
    [L^p_{ij}]_{1,1} &= |J_\pm|+|J_{\pm\pm}|\\
    [L^p_{ij}]_{2,2} &=  \tfrac{1}{4}|J_{zz}| + \tfrac{1}{2}|J_\pm|+\left(2+ \tfrac{1}{2}\textrm{Re}\g_{ij}+\tfrac{1}{4}\textrm{Im} \g_{ij}\right)|J_{\pm\pm}| \\
    [L^p_{ij}]_{3,3} &=  \tfrac{1}{4}|J_{zz}| + \tfrac{1}{2}|J_\pm|+\left(2- \tfrac{1}{2}\textrm{Re}\g_{ij}+\tfrac{1}{4}\textrm{Im} \g_{ij}\right)|J_{\pm\pm}| \\
    [L^p_{ij}]_{2,3} = [L^p_{ij}]_{3,2} &= \left(1-\tfrac{1}{4}\textrm{Im} \g_{ij}\right)|J_{\pm\pm}| \\
    [L^p_{ij}]_{4,4} &= \tfrac{1}{2}|J_\pm| + \tfrac{3}{2}|J_{\pm\pm}|
\end{aligned}$ \\
\end{tabular}
\end{ruledtabular}
\end{table*}

\section{Self-Consistent Equations} \label{appgeneralsc}
\subsection{The Standard Approach and Its Issues} \label{naive}
We summarize the usual argument to derive the self-consistent equations. After mean-field decoupling, the Hamiltonian \eqref{eq:hamb} is \begin{align}
    H_\textrm{MF} &= -\sum_{\la ij\ra} (\boldsymbol{B}_{ij}^\dagger \xi_{ij} \boldsymbol{\Aa}_{ij} + \boldsymbol{\Aa}^\dagger_{ij} \xi_{ij} \boldsymbol{B}_{ij} - \boldsymbol{\Aa}^\dagger_{ij} \xi_{ij} \boldsymbol{\Aa}_{ij}) \label{eq:apphambmf}
\end{align} where $\boldsymbol{B}_{ij}$ and $\boldsymbol{\Aa}_{ij}$ are defined in section \ref{maindecouple}. One can find the self-consistent equations by treating all mean-field amplitudes $\boldsymbol{\Aa}_{ij}$ as independent and then minimizing the mean-field energy 
\begin{align}
    \la H_{\textrm{MF}}\ra &= -\sum_{\la ij\ra} (\boldsymbol{X}_{ij}^\dagger \xi_{ij} \boldsymbol{\Aa}_{ij} + \boldsymbol{\Aa}^\dagger_{ij} \xi_{ij} \boldsymbol{X}_{ij} - \boldsymbol{\Aa}^\dagger_{ij} \xi_{ij} \boldsymbol{\Aa}_{ij}).  \label{eq:2}
\end{align} For brevity, we have denoted the expectation value of the vector $\boldsymbol{B}$ by $\boldsymbol{X}$. Then,
\begin{subequations}
    \begin{align}
    \frac{\di \la H_\textrm{MF}\ra}{\di \boldsymbol{\Aa}^\dagger_{k\ell}} = 0 &\iff \xi_{k\ell} (\boldsymbol{X}_{k\ell} - \boldsymbol{\Aa}_{k\ell}) = 0\\
     &\iff \boldsymbol{X}_{k\ell} = \boldsymbol{\Aa}_{k\ell} \label{eq:sc}
    \end{align}     
\end{subequations}
provided $\xi_{k\ell}$ is invertible. We call Eq. \eqref{eq:sc} the ``naive self-consistent equations", which holds for all bonds $(k,\ell)$. In the notation of previous work such as \cite{Desrochers2022}, Eq. \eqref{eq:sc} reads \begin{align}
    \la \hat{A}_{k\ell} \ra &= A_{k\ell}.
\end{align} Once \eqref{eq:sc} is satisfied, we have \begin{align}
    \la H_{\textrm{MF}}\ra\big|_\textrm{consistent} &= -\sum_{\la ij\ra} \boldsymbol{\Aa}_{ij}^\dagger \xi_{ij} \boldsymbol{\Aa}_{ij}.\label{eq:hcons}
\end{align} 
Now, we would like to impose PSG constraints on the ansatz. As argued in the main text, this can modify the form of the self-consistent equations beacuse the symmetries of the ansatz mean that the amplitudes on each bond are no longer independent. If the ansatz respects the PSG, there are only a few independent mean-field amplitudes (say, the set of $A_{01}$ for all $\hat{A}$), and the remaining $A_{ij}$ can depend on them in nontrivial ways. Therefore, one should apply the PSG constraints to the ansatz first and then minimize the energy with respect to the independent mean-field amplitudes. It can happen that the two approaches are equivalent if the decoupling and PSG constraints are simple enough, as was the case in the previously studied dipolar-octupolar model \cite{Desrochers2022}. 

An analogy can be used to illustrate this point: consider the constrained optimization problem \begin{align*}
    \textrm{minimize } &f(x,y) = x^2+y^2 \\
    \textrm{subject to } &g(x,y) = y-x-1=0
\end{align*} where $x,y \in \R$. In this analogy, $f$ corresponds to $\la H_\textrm{MF}\ra$ and $g$ corresponds to the PSG constraints. The equations to be solved to determine the minimum correspond to the self-consistent equations. If we minimized $f$ first and then applied the constraint, we would find $(x,y) = (0,0)$, which does not satisfy the constraint. So we fail to find any solutions. The ``self-consistent equations" in this case are 
\begin{subequations}
\begin{align}
    \frac{\di f(x,y)}{\di x} &= 0 \label{eq:badsc1} \\ 
    \frac{\di f(x,y)}{\di y} &= 0. \label{eq:badsc2}
\end{align}     
\end{subequations}
On the other hand, if we applied the constraint first (by solving for $y$ as a function of $x$) and then minimized with respect to $x$, we would get \[f(x, y(x)) = x^2 + (x+1)^2\] which has a minimum at $x=-\frac{1}{2}$. Thus, the minimum occurs at $(x,y) = (-\frac{1}{2}, \frac{1}{2})$, which is the correct solution to the constrained optimization problem. The ``self-consistent equation" in this case is \begin{align}
    \frac{\di f(x,y(x))}{\di x} &= 0, \label{eq:goodsc}
\end{align} and we see that Eqs. \eqref{eq:badsc1} and \eqref{eq:badsc2} are not the same as \eqref{eq:goodsc}. This is to be expected because the first approach does not actually work in general, as the above example shows. However, if the constraint was instead $g(x,y) = y-x$, then the first approach still works. This is only because we got lucky with the specific constraint. Thus, we are led to suspect that the self-consistent equations are not always \eqref{eq:sc}. 
\subsection{PSG Constraints on Bond Operator Expectation Values} \label{psg}
In appendix \ref{naive}, we defined vectors of expectation values $\boldsymbol{X}$. Here, it is clearer to separate $\boldsymbol{X}$ into two halves, $\boldsymbol{X} = (X^h, X^p)$ where the $4$-component vectors $X^\g$ have components
\begin{subequations}
\begin{align}
    [X^h_{ij}]^\alpha &= \la\boldsymbol{B}^\dagger_i \sigma^\alpha \boldsymbol{B}_j\ra \\
    [X^p_{ij}]^\alpha &= \la\boldsymbol{B}_i (i \sigma^y \sigma^\alpha) \boldsymbol{B}_j\ra.
\end{align}     
\end{subequations}
We claim that the $X^\g$ transform identically to the $t^\g$ vectors. Transforming the Schwinger bosons with gauge-enriched space group transformation $\tilde{\Oo}$ gives 
\begin{subequations}
\begin{align}
    \tilde{\Oo}: [X^h_{ij}]^\alpha &\mapsto \la\boldsymbol{B}^\dagger_{\Oo i} U_\Oo \sigma^\alpha U^\dagger_{\Oo}  \boldsymbol{B}_{\Oo j}\ra \nonumber\\
    &= \left[\mathcal{R}^h_\Oo X^h_{ij}\right]^\alpha \\
    \tilde{\Oo}: [X^p_{ij}]^\alpha &\mapsto \la\boldsymbol{B}_{\Oo i} U^*_{\Oo} (i \sigma^y \sigma^\alpha) U^\dagger_\Oo \boldsymbol{B}_{\Oo j}\ra \nonumber\\
    &= \left[\mathcal{R}^p_\Oo X^p_{ij}\right]^\alpha 
\end{align}     
\end{subequations}
where the matrices $\mathcal{R}_\Oo^\g $ were defined in Eq. \eqref{eq:abcdtransform} and we used the linearity of expectation value. But by definition, we have 
\begin{subequations}
\begin{align}
    [X^h_{\Oo i,\Oo j}]^\alpha &= \la\boldsymbol{B}^\dagger_{\Oo i} \sigma^\alpha \boldsymbol{B}_{\Oo j}\ra \\
    [X^p_{\Oo i,\Oo j}]^\alpha &= \la\boldsymbol{B}_{\Oo i} (i \sigma^y \sigma^\alpha) \boldsymbol{B}_{\Oo j}\ra,
\end{align}     
\end{subequations}
which tells us that
\begin{align}
    X^\g_{\Oo i,\Oo j} &= \mathcal{R}^\g_\Oo X^\g_{ij}, 
\end{align} which is structurally the same as Eq. \eqref{eq:abcdtransform}. Equivalently, the transformation law is \begin{align}
    \boldsymbol{X}_{ij} &= R_{ij} \boldsymbol{X}_{01}. \label{eq:expectransform} 
\end{align}  
\subsection{The Role of the Decoupling} \label{decoupling}
The previous two subsections have been independent of any details of the mean-field decoupling; that is, the matrix $\xi_{ij}$ in Eq. \eqref{eq:hamb}. We found that the transformation of the $t^\g_{ij}$ and $X^\g_{ij}$ only depends on the PSG, and not the decoupling used. Here, we will see that the decoupling only affects how the mean-field amplitudes $A_{ij}$ transform. A general mean-field decoupling will relate $\boldsymbol{t}$ to $\boldsymbol{\Aa}$ in the following way: 
\begin{align}
    \boldsymbol{t}_{ij} &= \xi_{ij} \boldsymbol{\Aa}_{ij}^*
\end{align} where $\xi_{ij} \in \R^{8\times 8}$ is precisely the symmetric matrix introduced in Eq. \eqref{eq:hamb}. This follows from the definition of the $t^\g$ vectors and the bond operators. For example, the singlet hopping term $\xi^{\chi,\chi}_{ij} \chi^*_{ij} \hat{\chi}_{ij}$ in $H_\textrm{MF}$ corresponds to 
\begin{align}
    a^h_{ij} &= \xi^{\chi,\chi}_{ij} \chi^*_{ij}
\end{align} because 
\begin{align}
    \boldsymbol{B}_i^{\ \dagger} (a^h_{ij} \openone_{2\times 2}) \boldsymbol{B}_j &= a^h_{ij} \hat{\chi}_{ij}.
\end{align} Since we have specialized to Hamiltonians of the form \eqref{eq:pseudospinham}, each pseudospin bilinear $S^\alpha_i S^\beta_j$ contains two $b^\dagger$ operators and two $b$ operators. Thus, $\xi$ takes the block-diagonal form \begin{align}
    \xi_{ij} &= \begin{pmatrix}
        L^h_{ij} & \mathbf{0}_{4\times 4} \\ 
        \mathbf{0}_{4\times 4} & L^p_{ij}
    \end{pmatrix}, \label{eq:ximatrix}
\end{align} where $L^\g_{ij} \in \R^{4\times 4}$ are real symmetric matrices. They must be real-valued and symmetric to ensure that $H_\textrm{MF}$ is Hermitian. We can use the transformation properties of $\boldsymbol{t}_{ij}$ from appendix \ref{psg} to deduce the transformation properties of $\boldsymbol{\Aa}_{ij}$. We have that 
\begin{align}
    \xi_{ij} \boldsymbol{\Aa}_{ij}^* &= \boldsymbol{t}_{ij} \nonumber\\
    &= R_{ij} \boldsymbol{t}_{01} \nonumber\\
    &= R_{ij} \xi_{01} \boldsymbol{\Aa}_{01}^*,
\end{align}     
from which it follows that 
\begin{align}
    \boldsymbol{\Aa}_{ij} &= S_{ij} \boldsymbol{\Aa}_{01} \label{eq:amptransform}
\end{align} where 
\begin{align}
    S_{ij} &= \xi^{-1}_{ij} R^*_{ij} \xi_{01} \label{eq:sdef} \\
    &= \begin{pmatrix}
     [L^h_{ij}]^{-1} \mathcal{R}^h_{ij} L^h_{01} & \mathbf{0}_{4\times 4} \nonumber\\
        \mathbf{0}_{4\times 4} & [L^p_{ij}]^{-1} \mathcal{R}^p_{ij} L^p_{01}
    \end{pmatrix}.
\end{align} For the PSG classes we will study, the $\mathcal{R}^\g$ matrices are real-valued, so $R^*_{ij} = R_{ij}$. We see that $\boldsymbol{\Aa}_{ij}$ depends linearly on $\boldsymbol{\Aa}_{01}$.
\subsection{The Generalized Self-Consistent Equations} 
To derive the general self-consistent equations, we proceed as outlined in appendix \ref{naive}; write all amplitudes and expectation values as functions of the reference bonds, and then minimize with respect to the independent amplitudes: 
\begin{subequations}
\begin{align}
        0 &= \frac{\di \la H_\textrm{MF}\ra}{\di \boldsymbol{\Aa}_{01}^\dagger} \\
        &= \sum_{\la ij\ra} \frac{\di \boldsymbol{\Aa}^\dagger_{ij}}{\di \boldsymbol{\Aa}_{01}^\dagger} \xi_{ij} (\boldsymbol{X}_{ij} - \boldsymbol{\Aa}_{ij}) \\
        &= \sum_{\la ij\ra} S^\intercal_{ij} \xi_{ij} (\boldsymbol{X}_{ij} - \boldsymbol{\Aa}_{ij}) \\
        &= \sum_{\la ij\ra} S^\intercal_{ij} \xi_{ij} (R_{ij}\boldsymbol{X}_{01} - S_{ij}\boldsymbol{\Aa}_{01}) \\
        &= N_\textrm{uc} (V \boldsymbol{X}_{01} - W \boldsymbol{\Aa}_{01}) 
\end{align}     
\end{subequations}
The self-consistent equations are 
\begin{align}
        \boldsymbol{X}_{01} &= (V^{-1} W)\boldsymbol{\Aa}_{01} \label{eq:generalsc} 
    \end{align} 
and we can immediately read off the matrices $V$ and $W$: 
\begin{subequations}
    \begin{align}
        V &= \sum_{\la ij\ra\in\textrm{uc}} S^\intercal_{ij} \xi_{ij} R_{ij} \label{eq:vdef}\\
        W &= \sum_{\la ij\ra\in\textrm{uc}} S^\intercal_{ij} \xi_{ij} S_{ij}. \label{eq:wdef}
    \end{align}
\end{subequations}
Note that the sum is only over the $12$ bonds in one pyrochlore unit cell. We emphasize that the general self-consistent equations relate $\boldsymbol{\Aa}_{01}$ to $\boldsymbol{X}_{01}$ only, as the previous subsections \ref{psg} and \ref{decoupling} showed how every other bond is already determined. 

The expressions (\ref{eq:vdef}-\ref{eq:wdef}) can be simplified considerably by expanding out the definitions of all the terms. We have
\begin{align}
        S^\intercal_{ij} \xi_{ij} R_{ij} &= (\xi_{01})^\intercal R_{ij}^\intercal (\xi_{ij}^{-1})^\intercal \xi_{ij} R_{ij} \nonumber\\
        &= \xi_{01}
\end{align}     
using the fact that $L^\g$ (and therefore $\xi$) is symmetric and $\mathcal{R}^\g$ (and therefore $R$) is orthogonal. Thus, \begin{align}
    V &= 12 \xi_{01}
\end{align} 
where the factor of $12$ appears because of the sum over $12$ bonds in the pyrochlore unit cell. Doing the same with $W$, 
\begin{align}
    S_{ij}^\intercal \xi_{ij} S_{ij}  &= \xi_{01}^\intercal R_{ij}^\intercal S_{ij} \nonumber\\
    &= \xi_{01} R_{ij}^{-1} S_{ij},
\end{align}     
which gives 
\begin{equation}
    W = \xi_{01} \sum_{\la ij\ra\in\textrm{uc}} R_{ij}^{-1} S_{ij}
\end{equation} 
and therefore
\begin{equation}
    V^{-1} W = \frac{1}{12} \sum_{\la ij\ra\in\textrm{uc}} R^{-1}_{ij} S_{ij}. \label{eq:VW}
\end{equation}     
This form makes it clear that all the $\Z_2$ phase factors cancel because the $R_{ij}$ appears twice. Thus, the nontrivial terms in the self-consistent equations stem from the structure of the $R$ matrices (determined by PSG constraints) and $\xi$ matrices (determined by the choice of decoupling) in Eq. \eqref{eq:VW}. The off-diagonal terms in the $R$ matrices (table \ref{table:bond_constraints_full}) combined with the off-diagonal terms in the decoupling (table \ref{table:fulldecoupling}) make the generalized self-consistent equations nontrivial for the amplitudes $E^x, E^y, D^x$, and $D^y$. Explicitly evaluating Eq. \eqref{eq:VW} gives the generalized self-consistent equations
\begin{subequations}
\begin{align}
    \la \chi_{01}\ra &= \chi_{01} \\
    \begin{pmatrix}
        \la E^x_{01}\ra \\ \la E^y_{01}\ra
    \end{pmatrix} &= \mathcal{M}^E \begin{pmatrix} E^x_{01} \\ E^y_{01} \end{pmatrix}\\
    \la E^z_{01}\ra &= E^z_{01} \\
    \la \Delta_{01}\ra &= \Delta_{01} \\
    \begin{pmatrix}
        \la D^x_{01}\ra \\ \la D^y_{01}\ra
    \end{pmatrix} &= \mathcal{M}^D \begin{pmatrix} D^x_{01} \\ D^y_{01} \end{pmatrix}\\
    \la D^z_{01}\ra &= D^z_{01} 
\end{align}    
\end{subequations} where the $2\times2$ matrices $\mathcal{M}^E, \mathcal{M}^D$ are given by 
\begin{subequations}
\begin{align}
    \mathcal{M}^E &= \frac{1}{d^E}\begin{pmatrix}
        x^E & 0 \\ 0 & y^E
    \end{pmatrix} L^h_{01} \big|_{E^x,E^y} \\
    \mathcal{M}^D &= \frac{1}{d^D}\begin{pmatrix}
        x^D & 0 \\ 0 & y^D
    \end{pmatrix}  L^p_{01} \big|_{D^x,D^y}. 
\end{align}    
\end{subequations} The notation $L^h_{01} \big|_{E^x,E^y}$ means the middle $2\times 2$ submatrix of $L^h_{01}$, or equivalently the restriction of $L^h_{01}$ to the subspace spanned by $(E^x, E^y)$. The terms $x^E, x^D, y^E, y^D, d^E, d^D$ are homogeneous polynomials in $J_{zz}, J_\pm, J_{\pm\pm}$ which depend on the exact decoupling used (see table \ref{table:fulldecoupling}), but are in general very unwieldy. For example, in the case $J_{zz}<0$, we find 
\begin{align}
    x^E &= 2(488J_{\pm\pm}^3 + 244J_{\pm\pm}^2 |J_{zz}| + 39 J_{\pm\pm}|J_{zz}|^2 + 2|J_{zz}|^3) \\
    y^E &= x_1 + 4J_{\pm\pm}J_{zz}(8J_{\pm\pm} + |J_{zz}|)\\
    d^E &= |J_{zz}| (8 J_{\pm\pm} + |J_{zz}|)(61 J_{\pm\pm}^2 + 16 J_{\pm\pm} |J_{zz}| + |J_{zz}|^2).
\end{align}

When the generalized self-consistent equations \eqref{eq:generalsc} are satisfied, the mean-field energy simplifies to \begin{align}
    \la H_{MF} \ra\big|_\textrm{consistent} &= - \sum_{\la ij\ra\in\textrm{uc}} \boldsymbol{\Aa}^\dagger_{ij} \xi_{ij} \boldsymbol{\Aa}_{ij},
\end{align} which is exactly Eq. \eqref{eq:hcons} that we found via the ``naive self-consistent equations". We can show this from some more expanding of definitions and applying Eq. \eqref{eq:amptransform} several times: 
\begin{subequations}
    \begin{align}
    \boldsymbol{\Aa}^\dagger_{ij} \xi_{ij} \boldsymbol{X}_{ij} &= \boldsymbol{\Aa}^\dagger_{ij} \xi_{ij} R_{ij} \boldsymbol{X}_{01} \\
    &= \boldsymbol{\Aa}^\dagger_{ij} \xi_{ij} R_{ij} V^{-1} W \boldsymbol{\Aa}_{01} \\
    &= \boldsymbol{\Aa}^\dagger_{ij} \xi_{ij} R_{ij} \left(\frac{1}{12}\sum_{\la k\ell\ra\in\textrm{uc}} R_{k\ell}^{-1} S_{k\ell}\right)  \boldsymbol{\Aa}_{01} \\
    &= \frac{1}{12} \sum_{\la k\ell\ra\in\textrm{uc}} (\boldsymbol{\Aa}_{01}^\dagger \xi_{01}^\intercal) R_{k\ell}^{-1} \boldsymbol{\Aa}_{k\ell} \\
    &= \frac{1}{12} \sum_{\la k\ell\ra\in\textrm{uc}} (\boldsymbol{\Aa}_{01}^\dagger \xi_{01}^\intercal R_{k\ell}^\intercal) \boldsymbol{\Aa}_{k\ell} \\
    &= \frac{1}{12} \sum_{\la k\ell\ra\in\textrm{uc}} \boldsymbol{\Aa}_{k\ell}^\dagger \xi_{k\ell} \boldsymbol{\Aa}_{k\ell},
\end{align}     
\end{subequations}
which is remarkably independent of $i, j$. It follows that the sum over nearest neighbours in the unit cell produces a factor of $12$, leading to the desired result \eqref{eq:hcons}.

\section{Explicit Expressions for Mean-Field Calculations}
\subsection{Fourier Transformation of the Mean-Field Hamiltonian}
The Fourier transform of the spinon operator is \begin{align}
    b_{\rr_\mu, \alpha} &= \frac{1}{\sqrt{N_\textrm{uc}}} \sum_{\kk \in \textrm{BZ}} b_{\kk,\mu,\alpha} e^{i\kk \cdot \rr}, \label{eq:ft}
\end{align} where $N_\textrm{uc}$ is the number of unit cells, BZ denotes the first Brillouin zone, and we use the convention that the exponential factor only depends on the unit cell position $\rr$, not on the sublattice. In terms of the vector 
$\Psi_\kk = \begin{pmatrix}
    \boldsymbol{b}_{\kk,0}, & \cdots & \boldsymbol{b}_{\kk,3}, \boldsymbol{b}^\dagger_{-\kk,0}, & \cdots & \boldsymbol{b}^\dagger_{-\kk,3}
\end{pmatrix}^\intercal$, the mean-field Hamiltonian \eqref{eq:hammfu} becomes \begin{align}
    H_\textrm{MF} &= \sum_{\kk \in \textrm{BZ}} \Psi^\dagger_\kk \Ha(\kk) \Psi_\kk \nonumber \\
    &- 4 N_\textrm{uc} \lambda (1+\kappa) + N_\textrm{uc} E_0 \label{eq:hammfft}
\end{align} where $E_0$ is shorthand for the following sum over unit cell bonds:
\begin{align}
    E_0 &= \sum_{\la ij \ra \in \textrm{uc}} \boldsymbol{\Aa}^\dagger_{ij} \xi_{ij} \boldsymbol{\Aa}_{ij}.
\end{align}
The $16\times 16$ matrix $\Ha(\kk)$ takes the standard Bogoliubov form 
\begin{align}
    \Ha(\kk) &= \begin{pmatrix}
        H_h(\kk) & H_p(\kk) \\
        H^\dagger_p(\kk) & H^\intercal_h(-\kk),
    \end{pmatrix}
\end{align} where the $8\times8$ blocks $H_h$ and $H_p$ satisfy $H_h(\kk) = H^\dagger_h(\kk)$ and $H_p(\kk) = H_p^\intercal(-\kk)$. Therefore, it is enough to specify the upper triangular part of the $H_\g$ matrices. Explicitly, 
\begin{align}
H_\g(\kk) &= \begin{pmatrix}
    0 & u_{01} + u_{I(01)} & u_{02} + u_{I(02)} & u_{03} + u_{I(03)} \\
    & 0 & u_{12} + u_{I(12)} & u_{13} + u_{I(13)} \\
    & & 0 & u_{23} + u_{I(23)} \\
    & & & 0
    \end{pmatrix}  
\end{align}
where we have defined 
\begin{subequations}
    \begin{align}
        u_{\mu\nu} &= u^\g_{\boldsymbol{0}_\mu, \boldsymbol{0}_\nu} \\
        u_{I(\mu\nu)} &= u^\g_{\boldsymbol{0}_\mu, (-\ee_\nu)_\nu} \exp(-i \kk \cdot \ee_\nu )
    \end{align}
\end{subequations} in terms of the $u^\g_{ij}$ matrices introduced in Eqs. (\ref{eq:uhmatrix}-\ref{eq:upmatrix}).

\subsection{Bogoliubov Transformation}
To diagonalize the mean-field Hamiltonian \eqref{eq:hammfft}, we perform a Bogoliubov transformation 
\begin{equation}
    \Psi_\kk = P(\kk) \Gamma_\kk \label{eq:bogo}
\end{equation}
where $\Gamma_\kk = \begin{pmatrix}
    \vec{\g}_{\kk,1}, & \cdots & \vec{\g}_{\kk,4}, \vec{\g}^\dagger_{-\kk,1}, & \cdots & \vec{\g}^\dagger_{-\kk,4}
\end{pmatrix}^\intercal$ and 
\begin{subequations}
\begin{align}
    P^\dagger(\kk) \Ha(\kk) P(\kk) &= \Lambda(\kk) \\
    &= \textrm{diag}(\omega_{\kk, 1}, \ldots, \omega_{\kk, 8}, \omega_{-\kk, 1}, \ldots, \omega_{-\kk, 8}). \label{eq:bogo evals}
\end{align}    
\end{subequations}
In order to preserve the bosonic commutation relations, $P(\kk)$ must satisfy the para-unitary condition
\begin{equation}
    P^\dagger(\kk) J P(\kk) = J, \quad J = \sigma^z \otimes \openone_{8\times8}.
\end{equation} 
We determine $P(\kk)$ by Colpa's algorithm, with the additional step of reordering the eigenvalues to appear as in Eq. \eqref{eq:bogo evals} \cite{Colpa1978}. 

With this change of basis \eqref{eq:bogo}, Eq. \eqref{eq:hammfft} becomes 
\begin{align}
    \frac{H_\textrm{MF}}{N_\textrm{uc}} &= \frac{1}{N_\textrm{uc}}\sum_{\kk \in \textrm{BZ}} \sum_{j=1}^8 \omega_{\kk,j} \g_{\kk,j}^\dagger \g_{\kk, j} \nonumber\\
    &+ \frac{1}{2N_\textrm{uc}} \sum_{\kk \in \textrm{BZ}} \sum_{j=1}^8 \omega_{\kk,j} - 4 \lambda(1+\kappa) + E_0.
\end{align}
The ground state $|0\ra$ satisfies $\g_{\kk,j,\alpha}|0\ra = 0$ for all $\kk, j, \alpha$, so the mean-field energy per unit cell is 
\begin{align}
    \frac{\la H_\textrm{MF}\ra}{N_\textrm{uc}} &= \frac{1}{2N_\textrm{uc}} \sum_{\kk \in \textrm{BZ}} \sum_{j=1}^8 \omega_{\kk,j} - 4 \lambda(1+\kappa) + E_0.
\end{align}  
\subsection{Evaluating Ground State Expectation Values}
To solve the (generalized) self-consistent equations, we use an iterative algorithm. We start with a random ansatz $\boldsymbol{\Aa}_{i_0,j_0}^{(0)}$ and fix a real number $\delta$, which is used in the convergence criterion. Step $n$ in the algorithm generates a new ansatz $\boldsymbol{\Aa}_{i_0,j_0}^{(n)}$ and Lagrange multiplier $\lambda^{(n)}$, which depend on $\boldsymbol{\Aa}_{i_0,j_0}^{(n-1)}$ and $\lambda^{(n-1)}$. In detail, step $n$ consists of the following:
\begin{enumerate}
    \item Solve for the Lagrange multiplier $\lambda^{(n)}$ using the self-consistency equation $\la n_i\ra = \kappa$. This can be done by bisection on the interval $\lambda \in (|h_\textrm{max}|,5)$ where $h_\textrm{max}$ is the largest eigenvalue of $H_\textrm{MF}$ over the first Brillouin zone. For the values of $\kappa, J_{zz}, J_\pm, J_{\pm\pm}$ considered in this work, it suffices to bound $\lambda$ above by $5$. 
    \item Using the values $\boldsymbol{\Aa}_{i_0,j_0}^{(n-1)},\lambda^{(n)}$ in $H_\textrm{MF}$, compute the ground state expectation values $\boldsymbol{X}_{i_0,j_0}^{(n)} = \la \boldsymbol{B}_{i_0,j_0}\ra$.
    \item Update $\boldsymbol{\Aa}_{i_0,j_0}^{(n)} = (W^{-1}V) \boldsymbol{X}_{i_0,j_0}^{(n)}$.
    \item Increment $n$ by one, return to step 1, and repeat until the desired convergence is achieved. Convergence is determined by the condition $\left|\boldsymbol{\Aa}_{i_0,j_0}^{(n)} - \boldsymbol{\Aa}_{i_0,j_0}^{(n-1)}\right| < \delta$.
\end{enumerate}

At each step of this algorithm, it is necessary to evaluate ground state expectation values $\la\boldsymbol{B}_{i_0,j_0} \ra$ for the reference bond $(i_0,j_0) =(\boldsymbol{0}_0,\boldsymbol{0}_1)$. For this calculation, it is convenient to write the Bogoliubov transformation matrix $P(\kk)$ in terms of $8\times8$ submatrices
\begin{align}
    P(\kk) &= \begin{pmatrix}
        P_{11}(\kk) & P_{12}(\kk) \\
        P_{21}(\kk) & P_{22}(\kk)
    \end{pmatrix},
\end{align} where the blocks satisfy 
\begin{equation}
    P_{22}(\kk) = P_{11}^*(-\kk), \quad P_{21}(\kk) = P_{12}^*(-\kk)
\end{equation}
due to the structure of the matrix $\Ha(\kk)$ \cite{Liu2019}. We can then write the Bogoliubov transformation \eqref{eq:bogo} as 
\begin{align}
    b_{\kk,\mu,\alpha} &= \sum_{\nu\beta} \Big([P_{11}(\kk)]_{(\mu\alpha),(\nu\beta)} \g_{\kk,\nu,\beta} \nonumber\\
    &+ [P_{12}(\kk)]_{(\mu\alpha),(\nu\beta)} \g^\dagger_{-\kk,\nu,\beta} \Big) \label{eq:bogo expand}
\end{align} where the composite sublattice-spin index $(\mu\alpha)$ represents a row or column index ranging from $1$ to $8$. In particular, if the spin index $\alpha$ is mapped to $\uparrow \ \mapsto 1$, $\downarrow \ \mapsto 2$, the composite index $(\mu\alpha)$ is equivalent to the usual matrix index $2\mu+\alpha$. Since the ground state $|0\ra$ is defined by $\g_{\kk,\mu,\alpha}|0\ra = 0$ for all $\kk, \mu, \alpha$, the only $\g$ bilinear that does not vanish immediately is 
\begin{align}
    \left\la \g_{\kk,\mu,\alpha} \g^\dagger_{\kk', \nu, \beta} \right\ra &= \left\la \delta_{\kk,\kk'} \delta_{\mu\nu} \delta_{\alpha\beta} + \g^\dagger_{\kk', \nu, \beta}\g_{\kk,\mu,\alpha} \right\ra \nonumber\\
    &=  \delta_{\kk,\kk'} \delta_{\mu\nu} \delta_{\alpha\beta}. \label{eq:groundexpec}
\end{align}

The hopping and pairing bond operators (\ref{eq:bond1})-(\ref{eq:bond4}) can be written compactly as follows; 
\begin{subequations}
    \begin{align}
    B_{\rr_\mu, \rr'_\nu}^{h,a} &= \sum_{\alpha\beta} b^\dagger_{\rr,\mu,\alpha} [\sigma^a]_{\alpha\beta} b_{\rr',\nu,\beta} \\
    B_{\rr_\mu, \rr'_\nu}^{p,a} &= \sum_{\alpha\beta} b_{\rr,\mu,\alpha} [i\sigma^y\sigma^a]_{\alpha\beta} b_{\rr',\nu,\beta} 
    \end{align}     
\end{subequations}
where the index $a \in \{0,1,2,3\}$ and $\sigma^0 = \openone_{2\times2}$. Performing the Fourier transformation \eqref{eq:ft}, the Bogoliubov transformation \eqref{eq:bogo expand}, and then making use of \eqref{eq:groundexpec} gives
\begin{subequations}
    \begin{align}
    \la B_{\rr_\mu, \rr'_\nu}^{h,a} \ra &= \frac{1}{N_\textrm{uc}} \sum_{\kk,\alpha\beta} \Big([P_{12}(\kk) P^\dagger_{12}(\kk)]_{(\nu\beta),(\mu\alpha)} \nonumber\\ 
    &\times [\sigma]^a_{\alpha\beta} e^{-i\kk \cdot(\rr-\rr')}\Big) \\
    \la B_{\rr_\mu, \rr'_\nu}^{p,a} \ra &= \frac{1}{N_\textrm{uc}} \sum_{\kk,\alpha\beta}  \Big([P_{11}(\kk) P^\dagger_{21}(\kk)]_{(\mu\alpha),(\nu\beta)} \nonumber\\
    &\times [i\sigma^y\sigma^a]_{\alpha\beta}e^{i\kk \cdot(\rr-\rr')}\Big).
    \end{align}    
\end{subequations}
\end{document}